\newcommand{\diracslash}[1]{#1\llap{/\kern2pt}}

\newcommand{\be}{\begin{equation}}
\newcommand{\ee}{\end{equation}}
\newcommand{\bea}{\begin{eqnarray}}
\newcommand{\eea}{\end{eqnarray}}
\newcommand{\ba}[1]{\begin{array}{#1}}
\newcommand{\ea}{\end{array}}

\newcommand{\bt}{\begin{tabular}}
\newcommand{\et}{\end{tabular}}

\newcommand{\beas}{\begin{eqnarray*}}
\newcommand{\eeas}{\end{eqnarray*}}

\documentclass[preprint,prd,aps,floats,nofootinbib,floatfix]{revtex4}

\DeclareSymbolFont{rsfs}{U}{rsfs}{m}{n}
\DeclareSymbolFontAlphabet{\mathrsfs}{rsfs}

\usepackage{graphicx}
\usepackage{multirow}
\usepackage{graphicx,epstopdf}

\usepackage{graphicx}
\usepackage{float}

\usepackage[utf8]{inputenc}
\usepackage[english]{babel}
\usepackage{hyperref}
\hypersetup{
    colorlinks=true,
    linkcolor=blue,
    filecolor=magenta,     
    citecolor=blue, 
    urlcolor=blue,
}
 
\usepackage{cleveref}

\begin{document}

\title{$J/\psi$ and $\eta_c$ in asymmetric hot  magnetized nuclear matter}

 \author{Rajesh Kumar}
\email{rajesh.sism@gmail.com}
\affiliation{Department of Physics, Dr. B R Ambedkar National Institute of Technology Jalandhar, 
 Jalandhar -- 144011,Punjab, India}
\author{Arvind Kumar}
\email{iitd.arvind@gmail.com, kumara@nitj.ac.in}
\affiliation{Department of Physics, Dr. B R Ambedkar National Institute of Technology Jalandhar, 
 Jalandhar -- 144011,Punjab, India}

\def\be{\begin{equation}}
\def\ee{\end{equation}}
\def\bearr{\begin{eqnarray}}
\def\eearr{\end{eqnarray}}
\def\zbf#1{{\bf {#1}}}
\def\bfm#1{\mbox{\boldmath $#1$}}
\def\hf{\frac{1}{2}}
\def\kp{\zbf k+\frac{\zbf q}{2}}
\def\km{-\zbf k+\frac{\zbf q}{2}}
\def\hwo{\hat\omega_1}
\def\hwt{\hat\omega_2}

\begin{abstract}
We investigate the mass-shift of vector channel $J/\psi$ and pseudoscalar channel $\eta_c$ in  strongly magnetized asymmetric nuclear medium at finite temperature using the conjunction of chiral $SU(3)$ model with the QCD sum rules. The magnetic field dependence of  scalar gluon condensate $\left\langle \frac{\alpha_{s}}{\pi} G^a_{\mu\nu} {G^a}^{\mu\nu} 
\right\rangle$ as well as  the twist-2 gluon condensate $\left\langle  \frac{\alpha_{s}}{\pi} G^a_{\mu\sigma}
{{G^a}_\nu}^{\sigma} \right\rangle $ calculated from  chiral SU($3$) model, are implemented  in QCD sum rules to calculate the magnetic field dependence of $J/\psi$ and $\eta_c$ meson masses. 
The effects of constant external magnetic field at finite density and temperature of the medium are found to be appreciable in symmetric and asymmetric nuclear matter. The results of the present investigation may be helpful to understand the experimental observables arising from the Compressed Baryonic Matter (CBM) produced in asymmetric non-central heavy ion collision experiments.

\end{abstract}

\maketitle

\maketitle

\section{Introduction}
\label{intro}
The investigation of the properties of hadrons in hot and dense matter under the effect of strong external magnetic field is a challenging area of research in non-perturbative QCD regime and of very importance from non-central Heavy-Ion Collisions (HICs) point of view which aim at understanding strong interaction physics. Due to highly asymmetric nature (different numbers of neutrons and protons) of heavy-ions, it is very imperative to include the isospin asymmetry while studying the properties of mesons in nuclear matter at finite temperature. In RHIC and LHC experiments, strongly interacting matter is produced at high temperature and very low baryon density. In addition, in non-central HICs, huge magnetic fields are believed to be produced and are estimated to be of the order of $eB\sim {15{m_\pi}^2}$ ($5\times 10^{19}$ gauss) at Large Hadron Collider (LHC) CERN and $eB\sim {2{m_\pi}^2}$ ($6.5\times 10^{18}$ gauss) at Relativistic Heavy Ion Collider (RHIC) BNL \cite{Kharzeev2008,Fukushima2008,Skokov2009}. The external magnetic field modifies the  QCD vacuum properties such as gluon condensates \cite{Bali2013,Ozaki2014} and quark condensates \cite{Schramm1992,Hong1996,Hong1998,Semenoff1999,Bali2012,Leung1996}.  Therefore, the study of the effect of magnetic field on the properties of mesons or hadronic medium has acquired a lot of attention recently \cite{Cho2014,Reddy2018,Gubler2016,Cho2015}.  With the construction of CBM experimental Facility for Antiproton and Ion Research (FAIR)and Nuclotron-based Ion Collider Facility (NICA) at Dubna, Russia, we expect significant progress to understand the in-medium modifications of hadrons at high baryonic density and moderate temperature.

 Strong magnetic fields also exist in the astrophysical compact objects namely magnetars. Magnetars are the neutron stars having very strong magnetic field \cite{Rabhi2011,Duncan1992}. It is suggested that the soft gamma-ray repeaters (SGRs) and anomalous X-ray pulsars (AXPs) reside in the category of magnetars \cite{Kouvellioton1998,Mareghetti1995}. The magnetic field at the interior and exterior of magnetars are believed to be as large as $ \sim 10^{18}$ gauss and $\sim 10^{14}$ gauss \cite{Shapiro1983}, respectively. The huge magnetic field in the centre is suggested to be due to the  dense quark matter magnetic core \cite{Ferrario2006,Iwazaki2005} and  has been studied substantially in literature \cite{Broderick2000,Cardall2001,Bandyopadhyay1997,
Mao2003,Rabhi2008}. Also, note that the magnetic field produced in the heavy-ion collisions is much above the field produced in magnetars and is comparable to QCD scale
and hence, will play an important role in understanding the QCD phase diagram.
Such strong magnetic field leads to many interesting physical phenomenon, for example, chiral magnetic effect \cite{Kharzeev2013,Fukushima2008,Vilenkin1980,Burnier2011}, chiral separation effect \cite{Kharzeev2013,Son2004,Son2007,Kharzeev2007}, heavy quark transport phenomenon \cite{Rapp2008}, magnetic catalysis of chiral symmetry breaking \cite{Kharzeev2013}, inverse magnetic catalysis \cite{Kharzeev2013}, medium modification of heavy flavoured mesons \cite{Gubler2016,Reddy2018,Cho2015},
magnetic inhibition at finite temperature and density \cite{Gusynin1994}.

Different theoretical models have been built to understand the properties of hadrons in the non-perturbative regime. 
Lattice QCD is the theoretical approach to study the hadron properties, and it is formulated on grid points of space and time \cite{Wilson1974}. This is applicable in the regime of low densities and high temperature, hence relevant for Relativistic Heavy Ion Collider (RHIC) and Large Hadron Collider (LHC)  experiments. This model is not applicable at high baryonic density due to sign problem \cite{Muroya2003,Bloch2009,Fukushima2007}. For finite baryonic density and temperature, phenomenological models based on basic QCD properties, for example, chiral symmetry, confinement properties, broken scale invariance, and trace anomaly properties are constructed. 
Walecka
model \cite{Walecka1974}, Nambu-Jona-Lasinio (NJL) model \cite{Nambu1961}, the Polyakov loop extended  (PNJL) model \cite{Fukushima2004,Kashiwa2008,Ghosh2015}, Chiral $SU(3)$ model \cite{Papazoglou1999,Mishra2004a,Mishra2009,Kumar2010,Kumar2011}, Coupled Channel approach \cite{Tolos2004,Tolos2006,Tolos2008,Hofmann2005}, Quark Meson Coupling (QMC) model \cite{Guichon1988,Hong2001,Tsushima1999,Sibirtsev1999,Saito1994,Panda1997},  Polyakov Quark Meson (PQM) model \cite{Chatterjee2012,Schaefer2010}, QCD sum rules \cite{Reinders1981,Hayashigaki2000,Hilger2009} are among the various approaches  which are used to explore the properties of mesons at finite density and temperature.

In the present paper, we will study the properties of ground state S-wave charmonium, $J/\psi$ and $\eta_c$, at finite density and temperature under strong external magnetic field using QCD sum rules and chiral $SU(3)$ model. QCD sum rules are used to evaluate hadronic properties, such as mass in term of condensates to get valuable information about the structure of QCD regime \cite{Reinders1981,Reinders1985,Klingl1997}. The mass modifications of heavy quarkonia, notably charmonium, and bottomonium have been investigated using QCD sum rules \cite{Cho2015,Cho2014,Kumar2010,Klingl1999,Morita2012,Kim2001,Suzuki2013,Mishra2014}. This model remarkably predicted the mass of $\eta_c$  in mixed states \cite{Shifman1978}. Also, within QCD sum rules, the in-medium mass and decay width of open charm mesons have been calculated \cite{Chhabra2017}. The chiral $SU(3)$ model has been used vastly to study the nuclear matter \cite{Kumar2010}, kaons and antikaons in nuclear and hyperonic matter \cite{Mishra2004}, finite nuclei \cite{Papazoglou1999} and  in-medium properties of vector mesons \cite{Zschiesche2004,Mishra2004}. This model is also generalized to $SU(4)$ to investigate the in-medium properties of D mesons \cite{Mishra2004a,Mishra2009,Kumar2010,Kumar2011}. To study the in-medium mass
of $J/\psi$ and $\eta_c$ mesons, we consider the contributions of 
the scalar gluon condensates, $\left\langle \frac{\alpha_{s}}{\pi}
G^a_{\mu\nu} {G^a}^{\mu\nu} \right\rangle$ and twist-2 tensorial gluon 
operator, $\left\langle  \frac{\alpha_{s}}{\pi} G^a_{\mu\sigma}
{{G^a}_{\nu}}^{\sigma} \right\rangle $ up to dimension four \cite{Klingl1997}. 
The scalar gluon condensate as well as the twist-2 gluon operator
in the asymmetric nuclear medium under the effect of finite magnetic field are calculated from chiral $SU(3)$ model \cite{Papazoglou1999} as will be discussed in detail in the next section.

Interest in the in-medium properties of charmonium was ignited when Matsui and Satz proposed that the decrease in the yield of $J/\psi$ state in HICs due to color screening effect should be considered as a probe of the production of Quark Gluon Plasma (QGP) \cite{Matsui1986}. Imperative results in favor of $J/\psi$ suppression were observed at CERN SPS and RHIC experiment \cite{Alessandro2005,Arnaldi2007,Adare2007}.
The statistical recombination of primordially produced charm 
quark pairs may lead to the increase in the yield of $J/\psi$ mesons and this picture is more important at LHC energies \cite{Ferrerio2014,Grandchamp2002,Andronic2003}. Charmonium is a bound state of charm and an anti-charm quark. The in-medium properties of charmonia are modified in the nuclear matter through gluon condensates \cite{Kumar2010,Lee2003}. The gluon condensates modify feebly, hence the obtained mass-shift of lowest charmonium states, $J/\psi$ and $\eta_c$  is very small in nuclear medium \cite{Klingl1999}. In  Ref. \cite{Lee2003}, the in-medium mass of charmonium states have been studied under linear density approximation in the nuclear medium using QCD second order stark effect. A significant mass =-shift is observed for the excited charmonium states ($\psi(3686)$ and $\psi(3770)$), but a small mass-shift for the ground state $J/\psi(3097)$. In Ref.  \cite{Morita2012}, mass shift of  $J/\psi$ and $\eta_c$ is studied using QCD sum rules through the medium modifications of gluon condensates in linear density approximation.  These studies are applicable up to nuclear saturation density only.
In Ref. \cite{Kumar2011} and \cite{Kumar2010}, gluon condensates evaluated in non-linear chiral $SU(3)$ model, in terms of dilaton field, were used as input in QCD second order stark effect and in QCD sum rules, respectively, to evaluate the masses of charmonium at finite density and temperature. In Ref. \cite{Dominguez2010,Mocsy2007,Asakawa2004}, the effect of temperature on the $J/\psi$ and $\eta_c$ was also studied in deconfined phase and observed that the heavy quark bound states can survive in the deconfined plasma. In addition to this, the in-medium properties of excited charmonium states, ${\chi_c}_0$ and ${\chi_c}_1$ are also studied in the literature \cite{Song2009,Morita2012}.  Recently, in Ref. \cite{Cho2015}, authors have investigated the masses of lowest charmonium states $J/\psi$ and $\eta_c$ in strong magnetic field using QCD sum rules. In Ref. \cite{Cho2014}, magnetically induced mixing between $J/\psi$ and $\eta_c$ have been calculated. Also, chiral $SU(3)$ model along with QCD second order stark effect was extended in Ref. \cite{Jahan2018} to study the effect of external magnetic field on the masses of $J/\psi$,  $\psi(3686)$ and $\psi(3770)$ mesons at finite density and zero temperature. 
$D$, $B$, and $\rho$ mesons are also studied in the presence of strong magnetic field \cite{Reddy2018,Gubler2016,Machado2014,Ghosh2016}. Due to the attractive potential of $J/\psi$ and $\eta_c$, it is also predicted that these mesons can form a bound state with nucleons \cite{Krein2011,Brodsky1990}.

 We organize the paper as follows. In Section \ref{sec:2}, we will present chiral $SU(3)$ model to describe the hot asymmetric
nuclear matter in the presence of
an external magnetic field. The scalar  density
as well as the number density of the charged proton has contributions from the Landau energy levels, whereas, the uncharged neutrons do not have contributions from Landau energy levels. In section \ref{sec:3}, we describe the QCD sum rules to calculate the in-medium mass-shift of charmonium. We discuss the results of the present investigation in section \ref{sec:4}, and finally in section \ref{sec:5} summary of this work will be given. 

\section{ The hadronic chiral $SU_L(3) \times SU_R(3)$ model}
\label{sec:2}

We use an effective field theoretical approach to describe hadron-hadron interactions, based on the broken scale invariance                                              \cite{Papazoglou1999,Mishra2004,Mishra2004a} and non-linear realization of chiral symmetry \cite{Weinberg1968,Coleman1969,Bardeen1969} in the presence of external magnetic field at finite density and temperature. In this model, the hadron-hadron interactions are expressed in terms of the scalar fields $\sigma$, $\zeta$, $\delta$, $\chi$ and vector fields $\omega$ and $\rho$. The $\delta$ and $\rho$ fields are introduced to incorporate the effect of isospin asymmetry in hadronic medium. Also, the dilaton field $\chi$ is proposed to express the scale symmetry breaking, leading to a non-vanishing trace of the energy-momentum tensor \cite{Papazoglou1999}. In this model, we use mean field approximation, which is a non-perturbative
relativistic approach to solve approximately the nuclear many-body problem. In this approximation, all the meson fields are considered as classical fields, hence only the vector and scalar fields contribute to nucleon-meson interaction Lagrangian term, as expectation value of the other mesons is zero \cite{Mishra2004a}. The Lagrangian density of this model under mean-field approximation is given as
\be
{\cal L}_{chiral} = {\cal L}_{kin} + \sum_{ M =S,V}{\cal L}_{NM}
          + {\cal L}_{vec} + {\cal L}_0 + {\cal L}_{SB}.
\label{genlag} \ee 

Individually,
\begin{eqnarray}
{\cal L}_{NM} = - \sum_{i} \bar {\psi_i} 
\left[ m_{i}^{*} + g_{\omega i} \gamma_{0} \omega 
+ g_{\rho i} \gamma_{0} \rho \right] \psi_{i},
\end{eqnarray}
\begin{eqnarray}
 {\cal L} _{vec} & = & \frac {1}{2} \left( m_{\omega}^{2} \omega^{2} 
+ m_{\rho}^{2} \rho^{2} \right) 
\frac {\chi^{2}}{\chi_{0}^{2}}
+  g_4 (\omega ^4 +6\omega^2 \rho^2+\rho^4),
\end{eqnarray}
\begin{eqnarray}
{\cal L} _{0} & = & -\frac{1}{2} k_{0}\chi^{2} \left( \sigma^{2} + \zeta^{2} 
+ \delta^{2} \right) + k_{1} \left( \sigma^{2} + \zeta^{2} + \delta^{2} 
\right)^{2} \nonumber\\
&+& k_{2} \left( \frac {\sigma^{4}}{2} + \frac {\delta^{4}}{2} + 3 \sigma^{2} 
\delta^{2} + \zeta^{4} \right) 
+ k_{3}\chi\left( \sigma^{2} - \delta^{2} \right)\zeta \nonumber\\
&-& k_{4} \chi^{4} 
 -  \frac {1}{4} \chi^{4} {\rm {ln}} 
\frac{\chi^{4}}{\chi_{0}^{4}}
+ \frac {d}{3} \chi^{4} {\rm {ln}} \Bigg (\bigg( \frac {\left( \sigma^{2} 
- \delta^{2}\right) \zeta }{\sigma_{0}^{2} \zeta_{0}} \bigg) 
\bigg (\frac {\chi}{\chi_0}\bigg)^3 \Bigg ),
\label{lagscal}
\end{eqnarray}

and
\begin{eqnarray}
{\cal L} _{SB} =  -\left( \frac {\chi}{\chi_{0}}\right)^{2} 
\left[ m_{\pi}^{2} 
f_{\pi} \sigma
+ \big( \sqrt {2} m_{K}^{2}f_{K} - \frac {1}{\sqrt {2}} 
m_{\pi}^{2} f_{\pi} \big) \zeta \right].
\label{lsb}
\end{eqnarray}

In Eq.(\ref{genlag}), ${\cal L}_{kin}$ represents the kinetic energy term, ${\cal L}_{NM}$ is the nucleon-meson interaction term, 
 where $S$ and $V$ represents the spin-0 and spin-1 mesons, respectively. Here, the effective mass of nucleons is given as
 \begin{equation}
  m_{i}^{*} = -(g_{\sigma i}\sigma + g_{\zeta i}\zeta + g_{\delta i}\tau_3 \delta).
  \label{mneff}
\end{equation}  
In above, $g_{\sigma i}$, $g_{\zeta i}$ and $g_{\delta i}$ represent the coupling strengths of  nucleons ($i$=$p,n$) with  $\sigma$, $\zeta$ and $\delta$ fields respectively and $\tau_3$ is the third component of isospin. The term $ {\cal L}_{vec}$ of Eq.(\ref{genlag}) generates the mass of vector mesons through the interactions with scalar mesons and contains the quartic self-interaction terms, $ {\cal L}_{0}$ describes  the spontaneous chiral symmetry breaking, and  ${\cal L}_{SB} $ describes the explicit chiral symmetry breaking.

In grand canonical ensemble, the partition function for nuclear system is given by \cite{Zschiesche1997} 
\bea
\mathcal{Z} &=& \mathrm{Tr \,exp}[-\beta(\hat{\mathcal{H}}-\sum_{i=p, n}
\mu_i \hat{\mathcal{N}}_i)],
\label{pfunc}
\eea
where $\beta=1/T$ and $\hat{\mathcal{H}}, \mu_i $ and $\hat{\mathcal{N}}_i$ are the Hamiltonian density operator, chemical potential and number density operator, respectively. The thermodynamical potential, $\Omega$ at given temperature $T$ reads
\bea
\Omega(T,V,\mu)=-T\ln\mathcal{Z}.
\eea
By inserting mean field Hamiltonian density \cite{Zschiesche1997} in terms of Lagrangian density, the thermodynamic potential $\Omega$, per unit volume, $V$ in zero magnetic field can be expressed as

\begin{equation}
\label{thermo}
\frac{\Omega} {V}= -\frac{\gamma_i T}
{(2\pi)^3} \sum_{i = p\,, n\, }
\int d^3k\biggl\{{\rm ln}
\left( 1+e^{-\beta [ E^{\ast}_i(k) - \mu^{*}_{i}]}\right) \\
+ {\rm ln}\left( 1+e^{-\beta [ E^{\ast}_i(k)+\mu^{*}_{i} ]}
\right) \biggr\} -{\cal L}_{vec} - {\cal L}_0 - {\cal L}_{SB}-{\mathcal{V}}_{vac},   
\end{equation}
where the sum runs over neutron and proton, $\gamma_i$ is the 
spin degeneracy factor for nucleons,
$E^{\ast}_i(k)=\sqrt{k^2+m^{*}_{i}}$ and 
$ \mu^{*}_{i}=\mu_{i}-g_{\omega i}\omega-g_{\rho i}\tau_{3}\rho$ are the effective single particle
energy of nucleons and effective nucleon chemical potential, respectively. In addition, vacuum potential energy, ${\mathcal{V}}_{vac}$  is subtracted in order to get zero vacuum energy.

In the presence of magnetic field, the Lagrangian density given by Eq.(\ref{genlag}) modifies to

\begin{equation}
{\cal L}_{T}={\cal L}_{chiral}+{\cal L}_{mag},
\label{Tlag}
\end{equation}
where
\be 
{\cal L}_{mag}=-{\bar {\psi_i}}q_i 
\gamma_\mu A^\mu \psi_i
-\frac {1}{4} \kappa_i \mu_N {\bar {\psi_i}} \sigma ^{\mu \nu}F^{\mu \nu}
\psi_i
-\frac{1}{4} F^{\mu \nu} F_{\mu \nu}.
\label{lmag}
\ee

In above $\psi_i$ is a wave function corresponds to the $i^{th}$ nucleon and the second term represents the tensorial interaction
with the electromagnetic field tensor, $F_{\mu \nu}$. Also, $k_i$ and $\mu_N$ are the anomalous magnetic moment of  $i^{th}$ nucleon and nuclear magneton, given as $\mu_N=\frac{e}{2m_N}$, respectively, where 
$m_N$ is the vacuum mass of the nucleon.
We choose the magnetic field to be uniform and along the
$Z$-axis, and hence the vector potential $A^\mu =(0,0,Bx,0)$. In the following, we discuss the interaction of charged and uncharged particles with external magnetic field. 

\subsection{Charged Proton in Magnetic Field}
In the presence of uniform magnetic field, due to the charged nature of proton, Lorentz force comes into picture. Hence, the transverse momenta of
proton with an electric charge $q_p$ are confined to discrete Landau levels, $\nu$, with, $k_\perp^2 = 2 \nu |q_p| B$, where 
$\nu \geq 0$ is an integral quantum number \cite{Strickland2012}. Thus, the volume integral converts into line integral  

\begin{equation}
\int {d^3}k \rightarrow \frac{|q_p| B}{(2\pi)^2} 
\sum_n \int_{0}^\infty d k_\parallel \, ,
\label{lsumform}
\end{equation}

where $k_\parallel $ is the momenta along the direction of magnetic field. Also, the summation represents a sum over the discrete orbital angular momentum, $n$ of proton in the perpendicular plane. 
The orbital quantum number  is related to $\nu$ via
$\nu=n+\frac{1}{2}-\frac{q_p}{|q_p|}\frac{s}{2}=0, 1, 2, \ldots$. Here, the quantum 
number $s$ is $+1$ for spin up and $-1$ for spin down protons. In addition to this, the effective single particle proton energy also gets quantized \cite{Ternov1966} and is given by

\bea
\tilde E^{p}_{\nu, s}&=&\sqrt{\left(k^{p}_{\parallel}\right)^{2}+
\left(\sqrt{m^{* 2}_{p}+2\nu |q_{p}|B}-s\mu_{N}\kappa_{p}B \right)^{2}}.
\eea

Under these changes, the first term of Eq.(\ref{thermo}), for proton will be
\bea
\label{thermop}
&\frac{\Omega_p} {V}= -
\frac{ T|q_p| B}{(2\pi)^2} \Bigg[
\sum_{\nu=0}^{\nu_{max}^{(s=1)}} \int_{0}^\infty d k_\parallel \, \biggl\{{\rm ln}
\left( 1+e^{-\beta [ \tilde E^{p}_{\nu, s} - \mu^{*}_{p} ]}\right) 
+ {\rm ln}\left( 1+e^{-\beta [ \tilde E^{p}_{\nu, s}+\mu^{*}_{p} ]}
\right) \biggr\}\nonumber\\
&+
\sum_{\nu=1}^{\nu_{max}^{(s=-1)}} \int_{0}^\infty d k_\parallel \, \biggl\{{\rm ln}
\left( 1+e^{-\beta [ \tilde E^{p}_{\nu, s} - \mu^{*}_{p} ]}\right)+ {\rm ln}\left( 1+e^{-\beta [ \tilde E^{p}_{\nu, s}+\mu^{*}_{p} ]}
\right) \biggr\}\Bigg].  
\eea

\subsection{Uncharged Neutron in Magnetic Field}

For uncharged neutron, there is no Landau quantization in the presence of external magnetic field, hence $\int  \frac{d^3k}{(2\pi)^3} \,$ remains unchanged \cite{Strickland2012}. The first term of thermodynamic potential, $\Omega$, per unit volume given by Eq.(\ref{thermo}), for neutron, will be

\begin{equation}
\label{thermon}
\frac{\Omega_n} {V}= -\frac{T}
{(2\pi)^3} \sum_{s=\pm 1}
\int d^3k\biggl\{{\rm ln}
\left( 1+e^{-\beta [ \tilde E^{n}_{s} - \mu^{*}_{n} ]}\right) \\
+ {\rm ln}\left( 1+e^{-\beta [\tilde E^{n}_{s}+\mu^{*}_{n}]}
\right) \biggr\},   
\end{equation}
where $\tilde E^{n}_{s}$ is the effective single particle energy of neutron in the presence of magnetic field and is given by
\bea
\tilde E^{n}_{s}&=& \sqrt{\left(k^{n}_{\parallel}\right)^{2} +
\left(\sqrt{m^{* 2}_{n}+\left(k^{n}_{\bot}\right)^{2} }-s\mu_{N}\kappa_{n}B 
\right)^{2}}.
\eea

The net thermodynamical potential in the presence of external magnetic field can be written as

\begin{equation}
\label{thermonet}
\frac{\Omega} {V}= \frac{\Omega_p} {V}+\frac{\Omega_n} {V} -{\cal L}_{vec} - {\cal L}_0 - {\cal L}_{SB}-{\mathcal{V}}_{vac}. 
\end{equation}

Now, by minimizing the
thermodynamical potential $\Omega/V$ of the nuclear system, the coupled equations of motion of the non-strange meson field $\sigma$, the strange scalar meson field $\zeta$, the scalar iso-vector meson field $\delta$, the vector meson field $\omega$, the vector-isovector meson field $\rho$, and  the dilaton field $\chi$, are determined and given as

\begin{eqnarray}
&&\frac{\partial (\Omega/V)}{\partial \sigma}= k_{0}\chi^{2}\sigma-4k_{1}\left( \sigma^{2}+\zeta^{2}
+\delta^{2}\right)\sigma-2k_{2}\left( \sigma^{3}+3\sigma\delta^{2}\right)
-2k_{3}\chi\sigma\zeta \nonumber\\
&-&\frac{d}{3} \chi^{4} \bigg (\frac{2\sigma}{\sigma^{2}-\delta^{2}}\bigg )
+\left( \frac{\chi}{\chi_{0}}\right) ^{2}m_{\pi}^{2}f_{\pi}
-\sum g_{\sigma i}\rho_{i}^{s} = 0,
\label{sigma}
\end{eqnarray}
\begin{eqnarray}
&&\frac{\partial (\Omega/V)}{\partial \zeta}= k_{0}\chi^{2}\zeta-4k_{1}\left( \sigma^{2}+\zeta^{2}+\delta^{2}\right)
\zeta-4k_{2}\zeta^{3}-k_{3}\chi\left( \sigma^{2}-\delta^{2}\right)\nonumber\\
&-&\frac{d}{3}\frac{\chi^{4}}{\zeta}+\left(\frac{\chi}{\chi_{0}} \right)
^{2}\left[ \sqrt{2}m_{K}^{2}f_{K}-\frac{1}{\sqrt{2}} m_{\pi}^{2}f_{\pi}\right]
 -\sum g_{\zeta i}\rho_{i}^{s} = 0 ,
\label{zeta}
\end{eqnarray}
\begin{eqnarray}
&&\frac{\partial (\Omega/V)}{\partial \delta}=k_{0}\chi^{2}\delta-4k_{1}\left( \sigma^{2}+\zeta^{2}+\delta^{2}\right)
\delta-2k_{2}\left( \delta^{3}+3\sigma^{2}\delta\right) +2k_{3}\chi\delta
\zeta \nonumber\\
& + &  \frac{2}{3} d \chi^4 \left( \frac{\delta}{\sigma^{2}-\delta^{2}}\right)
-\sum g_{\delta i}\tau_3\rho_{i}^{s} = 0 ,
\label{delta}
\end{eqnarray}

\begin{eqnarray}
\frac{\partial (\Omega/V)}{\partial \omega}=\left (\frac{\chi}{\chi_{0}}\right) ^{2}m_{\omega}^{2}\omega+g_{4}\left(4{\omega}^{3}+12{\rho}^2{\omega}\right)-\sum g_{\omega i}\rho_{i}^{v} = 0 ,
\label{omega}
\end{eqnarray}

\begin{eqnarray}
\frac{\partial (\Omega/V)}{\partial \rho}=\left (\frac{\chi}{\chi_{0}}\right) ^{2}m_{\rho}^{2}\rho+g_{4}\left(4{\rho}^{3}+12{\omega}^2{\rho}\right)-\sum g_{\rho i}\tau_3\rho_{i}^{v} = 0 ,
\label{rho}
\end{eqnarray}
and
\begin{eqnarray}
\frac{\partial (\Omega/V)}{\partial \chi}=k_{0}\chi \left( \sigma^{2}+\zeta^{2}+\delta^{2}\right)-k_{3}
\left( \sigma^{2}-\delta^{2}\right)\zeta + \chi^{3}\left[1
+{\rm {ln}}\left( \frac{\chi^{4}}{\chi_{0}^{4}}\right)  \right]
+(4k_{4}-d)\chi^{3}
\nonumber\\
-\frac{4}{3} d \chi^{3} {\rm {ln}} \Bigg ( \bigg (\frac{\left( \sigma^{2}
-\delta^{2}\right) \zeta}{\sigma_{0}^{2}\zeta_{0}} \bigg )
\bigg (\frac{\chi}{\chi_0}\bigg)^3 \Bigg )+
\frac{2\chi}{\chi_{0}^{2}}\left[ m_{\pi}^{2}
f_{\pi}\sigma +\left(\sqrt{2}m_{K}^{2}f_{K}-\frac{1}{\sqrt{2}}
m_{\pi}^{2}f_{\pi} \right) \zeta\right] \nonumber\\
- \frac{\chi}{{\chi^2}_0}(m_{\omega}^{2} \omega^2+m_{\rho}^{2}\rho^2)  = 0 ,
\label{chi}
\end{eqnarray}

respectively.

In above, $m_\pi$, $m_K$ and $f_\pi$, $f_K$ denote the mass and decay constant of $\pi$, $K$ mesons, respectively and the other parameters $k_0, k_2$ and $k_4$ are fitted to reproduce the vacuum mass of $\sigma$, $\zeta$ and $\chi$ meson and the remaining constants  $k_1$ is fixed to produce the effective nucleon mass at saturation density around 0.$65 m_N$ and $k_3$ is the constraint by $\eta$ and $\eta^\prime$ masses. Furthermore, $\rho^{s}_{i}$ and $\rho^{v}_{i}$ represent the scalar and vector/number densities of $i^{th}$ nucleon respectively.

 In the presence of the magnetic field, the vector density as well as
the scalar density of proton can be extracted from Eqs.(\ref{omega}) and (\ref{sigma})  and given as \cite{Broderick2000,Broderick2002}

\begin{eqnarray}
\rho^{v}_{p}=\frac{|q_{p}|B}{2\pi^2} \Bigg [ 
\sum_{\nu=0}^{\nu_{max}^{(s=1)}} \int^{\infty}_{0}
dk^p_{\parallel}\left( f^p_{k,\nu, s}-\bar{f}^p_{k,\nu, s}\right)
+\sum_{\nu=1}^{\nu_{max}^{(s=-1)}} \int^{\infty}_{0}
dk^p_{\parallel}\left( f^p_{k,\nu, s}-\bar{f}^p_{k,\nu, s}\right) 
\Bigg],
\label{rhovp}
\end{eqnarray}
and
\begin{equation}
\rho^{s}_{p}=\frac{|q_{p}|Bm^{*}_{p}}{2\pi^2} \Bigg [ 
\sum_{\nu=0}^{\nu_{max}^{(s=1)}}\int^{\infty}_{0}\frac{dk^p_{\parallel}}{\sqrt{(k^{p}_{\parallel})^2
+(\bar m_{p})^2}}\left( f^p_{k,\nu, s}+\bar{f}^p_{k, \nu, s}\right)
+\sum_{\nu=1}^{\nu_{max}^{(s=-1)}} \int^{\infty}_{0}\frac{dk^p_{\parallel}}{\sqrt{(k^{p}_{\parallel})^2
+(\bar m_{p})^2}}\left( f^p_{k,\nu, s}+\bar{f}^p_{k, \nu, s}\right)
\Bigg],
\label{rhosp}
\end{equation}

respectively, where $\bar m_{p}$ is the effective mass under the effect of magnetic field, defined as
\bea
\bar m_{p}=\sqrt{m^{* 2}_{p}+2\nu |q_{p}|B}-s\mu_{N}\kappa_{p}B.
\label{mc}
\eea

Similarly, for the uncharged neutron, the number  and scalar densities are given by \cite{Broderick2000,Broderick2002}

\bea
\rho^{v}_{n}&=&\frac{1}{2\pi^{2}}\sum_{s=\pm 1}\int^{\infty}_{0}k^{n}_{\bot}
dk^{n}_{\bot} \int^{\infty}_{0}\, dk^{n}_{\parallel}
\left( f^n_{k, s}-\bar{f}^n_{k, s}\right), 
\label{rhovn} 
\eea
and
\bea
\rho^{s}_{n}&=&\frac{1}{2\pi^{2}}\sum_{s=\pm 1}\int^{\infty}_{0}
k^{n}_{\bot}dk^{n}_{\bot}\left(1-\frac{s\mu_{N}\kappa_{n}B}
{\sqrt{m^{* 2}_{n}+\left(k^{n}_{\bot}\right)^{2}}} \right)  
\int^{\infty}_{0}\, dk^{n}_{\parallel}
\frac{m^*_n}{\tilde E^{n}_{s}}\left(
f^n_{k, s}+\bar{f}^n_{k, s}\right),
\label{rhosn} 
\eea
respectively. In above, ${f}^p_{k, \nu, s}$, $\bar{f}^p_{k, \nu, s}$,  ${f}^n_{k, s}$ and $\bar{f}^n_{k, s}$ represent the  finite temperature distribution functions for particles and antiparticles  for proton and neutron, and given as
\bea
f^p_{k,\nu, s} &=& \frac{1}{1+\exp\left[\beta(\tilde E^p_{\nu, s} 
-\mu^{*}_{p}) \right]}, \qquad
\bar{f}^p_{k,\nu, s} = \frac{1}{1+\exp\left[\beta(\tilde E^p_{\nu, s} 
+\mu^{*}_{p} )\right]},
\label{dfp}
\eea
\bea
f^n_{k, s} &=& \frac{1}{1+\exp\left[\beta(\tilde E^n_{s} 
-\mu^{*}_{n}) \right]}, \qquad
\bar{f}^n_{k, s} = \frac{1}{1+\exp\left[\beta(\tilde E^n_{s} 
+\mu^{*}_{n} )\right]}.
\label{dfn}
\eea

If we neglect the effect of magnetic field in Eq.(\ref{Tlag}), the vector and scalar densities of the nucleons are 
modified and given as 

\begin{eqnarray}
\rho_{i}^{v} = \gamma_{i}\int\frac{d^{3}k}{(2\pi)^{3}}  
\Bigg(\frac{1}{1+\exp\left[\beta(E^{\ast}_i(k) 
-\mu^{*}_{i}) \right]}-\frac{1}{1+\exp\left[\beta(E^{\ast}_i(k)
+\mu^{*}_{i}) \right]}
\Bigg),
\label{rhov0}
\end{eqnarray}

 and

\begin{eqnarray}
\rho_{i}^{s} = \gamma_{i}\int\frac{d^{3}k}{(2\pi)^{3}} 
\frac{m_{i}^{*}}{E^{\ast}_i(k)} \Bigg(\frac{1}{1+\exp\left[\beta(E^{\ast}_i(k) 
-\mu^{*}_{i}) \right]}+\frac{1}{1+\exp\left[\beta(E^{\ast}_i(k)
+\mu^{*}_{i}) \right]}
\Bigg),
\label{rhos0}
\end{eqnarray}
respectively.

As discussed earlier, the effect of isospin asymmetry  is introduced by incorporating $\delta$ and $\rho$ fields having dependence on the isospin asymmetry parameter, $\eta=(\rho^{v}_{n} -{\rho^{v}_{p}} )/2\rho_N
$, which is the measure of the abundance of one isospin component over other. 

In the present investigation, gluon condensates will be used to calculate the effective mass of charmonia within QCD sum rules. The scalar gluon condensate $\left\langle \frac{\alpha_{s}}{\pi} G^a_{\mu\nu} {G^a}^{\mu\nu} 
\right\rangle$ and the twist-2 tensorial gluon operator, 
$\left\langle  \frac{\alpha_{s}}{\pi} G^a_{\mu\sigma}
{{G^a}_\nu}^{\sigma} \right\rangle $, given by Eqs.(\ref{chiglum})
and (\ref{g2approx}) respectively in the following are derived in terms of medium modified $\sigma$, $\zeta$, $\delta$ and dilaton field $\chi$ within chiral $SU(3)$ model as explained below.

Within this model, gluon condensates are obtained from the energy-momentum tensor in terms of $\chi$ field \cite{Kumar2010}. We start with energy-momentum tensor
\begin{eqnarray}
T_{\mu \nu}=(\partial _\mu \chi) 
\Bigg (\frac {\partial {{\cal L}_\chi}}
{\partial (\partial ^\nu \chi)}\Bigg )
- g_{\mu \nu} {\cal L}_\chi,
\label{energymom}
\end{eqnarray}
where ${\cal L}_\chi $ represents the Lagrangian density term, which is written in the model to incorporate scale breaking property of QCD and is given by
\begin{eqnarray}
{\cal L}_\chi & = & \frac {1}{2} (\partial _\mu \chi)(\partial ^\mu \chi)- k_4 \chi^4 - \frac{1}{4} \chi^{4} {\rm {ln}} 
\Bigg ( \frac{\chi^{4}} {\chi_{0}^{4}} \Bigg )
+ \frac {d}{3} \chi^{4} {\rm {ln}} \Bigg (\bigg( \frac {\left( \sigma^{2} 
- \delta^{2}\right) \zeta }{\sigma_{0}^{2} \zeta_{0}} \bigg) 
\bigg (\frac {\chi}{\chi_0}\bigg)^3 \Bigg ).
\label{lagchi}
\end{eqnarray}
To obtain the trace of above energy-momentum tensor, we multiply it by $g^{\mu \nu}$,

\begin{eqnarray}
T_{\mu}^{\mu} = (\partial _\mu \chi) \Bigg (\frac {\partial {{\cal L}_\chi}}
{\partial (\partial _\mu \chi)}\Bigg ) -4 {{\cal L}_\chi}=-(1-d)\chi^{4}. 
\label{traceemchi}
\end{eqnarray}

The energy-momentum tensor, $T_{\mu \nu}$ and its trace, $T_{\mu}^{\mu} $ in massless QCD   can be written as
\cite{Lee2009,Morita2008} 
\begin{eqnarray}
T_{\mu \nu}=-\Big (\frac{\pi}{\alpha_s}\Big)\Big (u_\mu u_\nu - 
\frac{g_{\mu \nu}}{4} \Big ) G_2
+ \frac {g_{\mu \nu}}{4} \frac{\beta_{QCD}}{2g} 
{G^a}_{\sigma \kappa} {G^a}^{\sigma \kappa},
\label{emqcd}
\end{eqnarray}

and

\begin{eqnarray}
T_{\mu}^{\mu} = \langle \frac{\beta_{QCD}}{2g} 
{G^a}_{\sigma \kappa} {G^a}^{\sigma \kappa} \rangle ,
\label{traceemqcd}
\end{eqnarray}
respectively. In Eq.(\ref{emqcd}), $G_2$ is twist-2 gluon operator.

By comparing Eq.(\ref{traceemchi}) with (\ref{traceemqcd}), scalar gluon condensate, $G_0$ can be written as
\begin{eqnarray}
G_0=\left\langle  \frac{\alpha_{s}}{\pi} {G^a}_{\mu\nu} {G^a}^{\mu\nu} 
\right\rangle =  \frac{8}{9} \Bigg [(1 - d) \chi^{4}
\Bigg ].
\label{chiglu}
\end{eqnarray}

If we take the effect of finite quark mass term \cite{Kumar2010}, the modified scalar gluon condensate, $G_0$ is expressed  as
\begin{eqnarray}
G_0=\left\langle  \frac{\alpha_{s}}{\pi} {G^a}_{\mu\nu} {G^a}^{\mu\nu} 
\right\rangle =  \frac{8}{9} \Bigg [(1 - d) \chi^{4}
+\left( \frac {\chi}{\chi_{0}}\right)^{2} 
\left( m_{\pi}^{2} f_{\pi} \sigma
+ \big( \sqrt {2} m_{k}^{2}f_{k} - \frac {1}{\sqrt {2}} 
m_{\pi}^{2} f_{\pi} \big) \zeta \right) \Bigg ].
\label{chiglum}
\end{eqnarray}

The comparison of energy-momentum tensor in chiral model with QCD \cite{Kumar2010}, gives the expression for twist-2 gluon operator and is given by

\begin{eqnarray}
G_2=\left\langle  \frac{\alpha_{s}}{\pi} G^a_{\mu\sigma}
{{G^a}_\nu}^{\sigma} \right\rangle &=& 
 \frac{\alpha_{s}}{\pi}\Bigg [-(1-d+4 k_4)(\chi^4-{\chi_0}^4)-\chi ^4 {\rm {ln}}
\Big (\frac{\chi^4}{{\chi_0}^4}\Big )\nonumber \\ 
& + & \frac {4}{3} d\chi^{4} {\rm {ln}} \Bigg (\bigg( \frac {\left( \sigma^{2} 
- \delta^{2}\right) \zeta }{\sigma_{0}^{2} \zeta_{0}} \bigg) 
\bigg (\frac {\chi}{\chi_0}\bigg)^3 \Bigg ) \Bigg ].
\label{g2approx}
\end{eqnarray}

The value of $d$ may be taken from QCD beta function, $\beta_{QCD}$  at the one loop level \cite{Papazoglou1999}, 
with $N_c$ colors and $N_f$ flavors, 
\be
\label{qcdbeta}
   \beta_{QCD}=-\frac{11 N_c g^3}{48 \pi^2} \left(1-\frac{2N_f}{11 N_c}\right)
   +{\cal O}(g^5) . 
\ee                               
Here, the first term in parentheses arises from the 
self-interaction of the gluons and the second term is proportional to $N_f$, 
is the contribution from quark pairs. This equation  suggests 
the value $d$=$6/33$ for three flavors and three colors.

\section{In-medium masses of $J/\psi$ and $\eta_c$ meson in external magnetic field}
\label{sec:3}

In the present investigation, to study the effective mass of charmonium, we use QCD sum rules, which is a non-perturbative mechanism to study the confining nature of QCD \cite{Reinders1985,Reinders1981,Klingl1999}. In this approach, effective mass of mesons is obtained from the moment derived from the operator product expansion (OPE) of correlation function \cite{Reinders1981}.

 We start from time ordered current-current correlation function of two heavy quark currents 
 \cite{Reinders1985}.
\begin{equation}
\Pi^{\, J}(q) = i \int d^4x e^{i q \cdot x} \langle T[j^{\, c}(x) j^{\, c \dag}(0)] \rangle . \label{correlation function}
\end{equation}
 Here, $q=( \omega, \vec{q})$, represents four momentum vector and symbol  $c$ stands for the pseudoscalar $(P)$ and vector $(V)$ mesons. Each current is defined as $j^{\, P}=\bar{\psi_q} \gamma_5 \psi_q$ and $ j^{\, V}_\mu = \bar{\psi_q} \gamma_\mu \psi_q$ with $\psi_q$ being the quark operator. 

We can render the current-current correlation in the region of positive and large 
$Q^2=\vec{q}^{\, 2}-\omega^2$ through an OPE. Hence, left-hand side of above equation can be written as \cite{Wilson1969}
  \begin{eqnarray}
\label{eq2}
\Pi (q)=\sum_n W_{n}(q) \, 
\langle \hat O_{n} \rangle.
\end{eqnarray}
In above, the $\hat O_n$ are operators of  dimension $n$, and $W_{n}$ are the perturbative Wilson coefficients.

By expressing correlation function as polarization function $\tilde{\Pi}^J$, which reduces to vacuum polarization function in the limit $\rho_N \rightarrow$ 0 \cite{Klingl1999}, the  $n^{th}$ moment can be expressed as
\begin{eqnarray}
\label{dispersion}
M_n^J &\equiv& \left. { 1 \over n!} \left( {d \over d \omega^2} \right)^n
\tilde{\Pi}^{J}(\omega^2) \right|_{\omega^2=-Q_0^2}, \nonumber \\ &&=
\frac{1}{\pi}\int_{4m_{c}^2}^{\infty}\frac{\mbox{Im}
\tilde{\Pi}^{J}(s)}{(s+Q_0^2)^{n+1}}ds,
\end{eqnarray}
at a fixed $Q_0^2=4m_c^2 \xi$.

Using operator product expansion, the moment 
$M_{n}^{J}$ can be written as \cite{Klingl1997}
\begin{equation}
M_{n}^{J} (\xi) = A_{n}^{J} (\xi) \left[  1 + a_{n}^{J} (\xi) \alpha_{s} 
+ b_{n}^{J} (\xi) \phi_{b} + c_{n}^{J} (\xi) \phi_{c} \right],
\label{moment}  
\end{equation}
where $A_n^J(\xi)$, $a_n^J(\xi)$, $b_n^J (\xi)$ and $c_n^J (\xi)$
are the Wilson coefficients \cite{Reinders1981,Klingl1997} and $\xi$ is the 
normalization scale. The common factor $A_{n}^{J}$ results 
from the bare loop diagram. The coefficients $a_{n}^{J}$ take into 
account perturbative radiative corrections, while the coefficients 
$b_{n}^{J}$ are associated with the scalar gluon condensate through
\begin{equation}
\phi_{b} = \frac{4 \pi^{2}}{9} \frac{\left\langle \frac{\alpha_{s}}{\pi} 
G^a_{\mu \nu} {G^a}^{\mu \nu} \right\rangle }{(4 m_{c}^{2})^{2}}.
\label{phib} 
\end{equation}
As described earlier, the contribution of the scalar gluon condensate 
is incorporated through the $\sigma$, $\zeta$, $\delta$ and dilaton field $\chi$ within the chiral $SU(3)$ model. 
Using Eq.(\ref{chiglum}), the above equation can be rewritten
in terms of these fields as
\begin{equation}
\phi_{b} =  \frac{32 \pi^{2}}{81(4 m_{c}^{2})^{2}}  
\Bigg [ (1-d)\chi^4 
+\left( \frac {\chi}{\chi_{0}}\right)^{2} 
\left( m_{\pi}^{2} f_{\pi} \sigma
+ \big( \sqrt {2} m_{K}^{2}f_{K} - \frac {1}{\sqrt {2}} 
m_{\pi}^{2} f_{\pi} \big) \zeta \right) \Bigg ]. 
\label{phibglu}
\end{equation} 
The coefficients $c_n^J$ are associated with the 
value of $\phi_{c}$, which gives the contribution from twist-2 gluon 
operator, $\phi_c$ appearing in Eq.(\ref{moment}) is defined by
\begin{equation}
\phi_{c} =  \frac{4 \pi^{2}}{3(4 m_{c}^{2})^{2}}\left\langle  \frac{\alpha_{s}}{\pi} G^a_{\mu\sigma}
{{G^a}_\nu}^{\sigma} \right\rangle 
,  
\label{phic}
\end{equation}
where twist-2 gluon operator $\left\langle  \frac{\alpha_{s}}{\pi} G^a_{\mu\sigma}
{{G^a}_\nu}^{\sigma} \right\rangle $ 
 is given by Eq.(\ref{g2approx}). Explicitly in terms of $\sigma$, $\zeta$, $\delta$ and dilaton field $\chi$, $\phi_c$ will be,
 \begin{equation}
\phi_{c} =  \frac{4 \pi^{2}\alpha_s}{3(4 m_{c}^{2})^{2}\pi}\Bigg [-(1-d+4 k_4)(\chi^4-{\chi_0}^4)-\chi ^4 {\rm {ln}}
\Big (\frac{\chi^4}{{\chi_0}^4}\Big )\nonumber \\ 
 +\frac {4}{3} d\chi^{4} {\rm {ln}} \Bigg (\bigg( \frac {\left( \sigma^{2} 
- \delta^{2}\right) \zeta }{\sigma_{0}^{2} \zeta_{0}} \bigg) 
\bigg (\frac {\chi}{\chi_0}\bigg)^3 \Bigg ) \Bigg ]
.  
\label{phicglu}
\end{equation}

The $m_{c}$ and $\alpha_{s}$ parameters are the running charm quark mass 
and running coupling constant and are $\xi$ dependent \cite{Reinders1981}, and are given by
\begin{equation}
\frac{m_{c} (\xi)}{m_{c}} = 1 - \frac{\alpha_{s}}{\pi} \left\lbrace 
\frac{2 + \xi}{1 + \xi} {\rm {ln}}(2 + \xi) - 2 {\rm {ln}}2\right\rbrace ,
\label{mcz}
\end{equation}
where, $m_{c} \equiv m_{c} (p^{2} = -m_{c}^{2}) = 1.26$ GeV \cite{Reinders1985}, and
\begin{eqnarray}
\alpha_{s}\left( Q_{0}^{2} + 4 m_{c}^{2} \right)  & = &  
\alpha_{s}\left( 4 m_{c}^{2} \right)/\left( 1 + \frac{25}{12\pi} 
\alpha_{s}\left( 4 m_{c}^{2}\right)
{\rm {ln}} \frac{Q_{0}^{2} + 4 m_{c}^{2}}{4 m_{c}^{2}} \right),
\end{eqnarray}
with, $\alpha_{s}\left( 4 m_{c}^{2} \right) \simeq 0.3$ and 
$Q_{0}^{2} = 4 m_{c}^{2}\xi$ \cite{Reinders1981}.

The in-medium mass of the $J/\psi$ and $\eta_c$ charmonium states in terms of $M_{n}^{J}$ can be written as
\begin{equation}
{{m^*_C}}^{2} \simeq \frac{M_{n-1}^{J} (\xi)}{M_{n}^{J} (\xi)} - 4 m_{c}^{2} \xi.
\label{masscharm}
\end{equation}

In the next section, we shall discuss the results for 
the $J/\psi$ and $\eta_c$ meson mass modifications in asymmetric nuclear
matter in the presence of an external magnetic field, $B$, at finite temperature.
\section{Results and Discussions}
\label{sec:4}
The effects of external magnetic field in hot and dense nuclear matter on the properties of $J/\psi$ and $\eta_c$ in the present work are calculated through the in-medium behaviour of scalar gluon condensate $\left\langle \frac{\alpha_{s}}{\pi} G^a_{\mu\nu} {G^a}^{\mu\nu} 
\right\rangle$ and the twist-2 tensorial gluon operator, 
$\left\langle  \frac{\alpha_{s}}{\pi} G^a_{\mu\sigma}
{{G^a}_\nu}^{\sigma} \right\rangle $. We have divided our discussion in three sections. In Section A, we will discuss the effect of external magnetic field on in-medium behaviour of scalar fields $\sigma$, $\zeta$, $\delta$ and $\chi$, through which medium dependence of gluon condensates will be evaluated. The in-medium effects on gluon condensates are discussed in Section B. Finally, in Section C, medium modification of $J/\psi$ and $\eta_c$ meson will be  discussed.

\subsection{  Scalar Fields $\sigma$, $\zeta$, $\delta$, and $\chi$ in Hot Magnetized Nuclear Medium} 
\label{fields}

 \begin{table}
\begin{tabular}{|c|c|c|c|c|}
\hline 
$g_{\sigma N}$  & $g_{\zeta N }$  &  $g_{\delta N }$  &
$g_{\omega N}$ & $g_{\rho N}$ \\

\hline 
10.56 & -0.46 & 2.48 & 13.35 & 5.48  \\ 
\hline
\hline

\hline
$k_0$ & $k_1$ & $k_2$ & $k_3$ & $k_4$  \\ 
\hline 
2.53 & 1.35 & -4.77 & -2.77 & -0.218  \\ 
\hline
\hline

\hline 
$m_\pi $(MeV) &$ m_K$ (MeV) &$ f_\pi$ (MeV) & $f_K$(MeV) & $g_4$ \\ 
\hline 
139 & 498 & 93.29 & 122.14 & 79.91  \\ 
\hline
\hline

\hline 
$\sigma_0$ (MeV) & $\zeta_0$ (MeV) & $\chi_0$ (MeV) & $d$ & $\rho_0$ ($\text{fm}^{-3}$)  \\ 
\hline 
-93.29 & -106.8 & 409.8 & 0.064 & 0.15  \\ 
\hline

\end{tabular}
\caption{Values of various parameters.} \label{ccc}
\end{table} 

 By solving coupled system of non-linear equations ((\ref{sigma}) to (\ref{chi})) for  $\sigma$, $\zeta$, $\delta$, $\omega$, $\rho$ and $\chi$, temperature and density dependence of these fields in the presence of magnetic field is calculated. These equations have the dependence on vector density, $\rho^{v}_i$ and scalar density, $\rho^{s}_i$ of the nucleons, which further have magnetic field dependency as can be seen from Eqs.(\ref{rhovp}) to (\ref{rhosn}). Various parameters used in the present work are listed in table \ref{ccc}.

\begin{figure}[h]
\includegraphics[width=16cm,height=16cm]{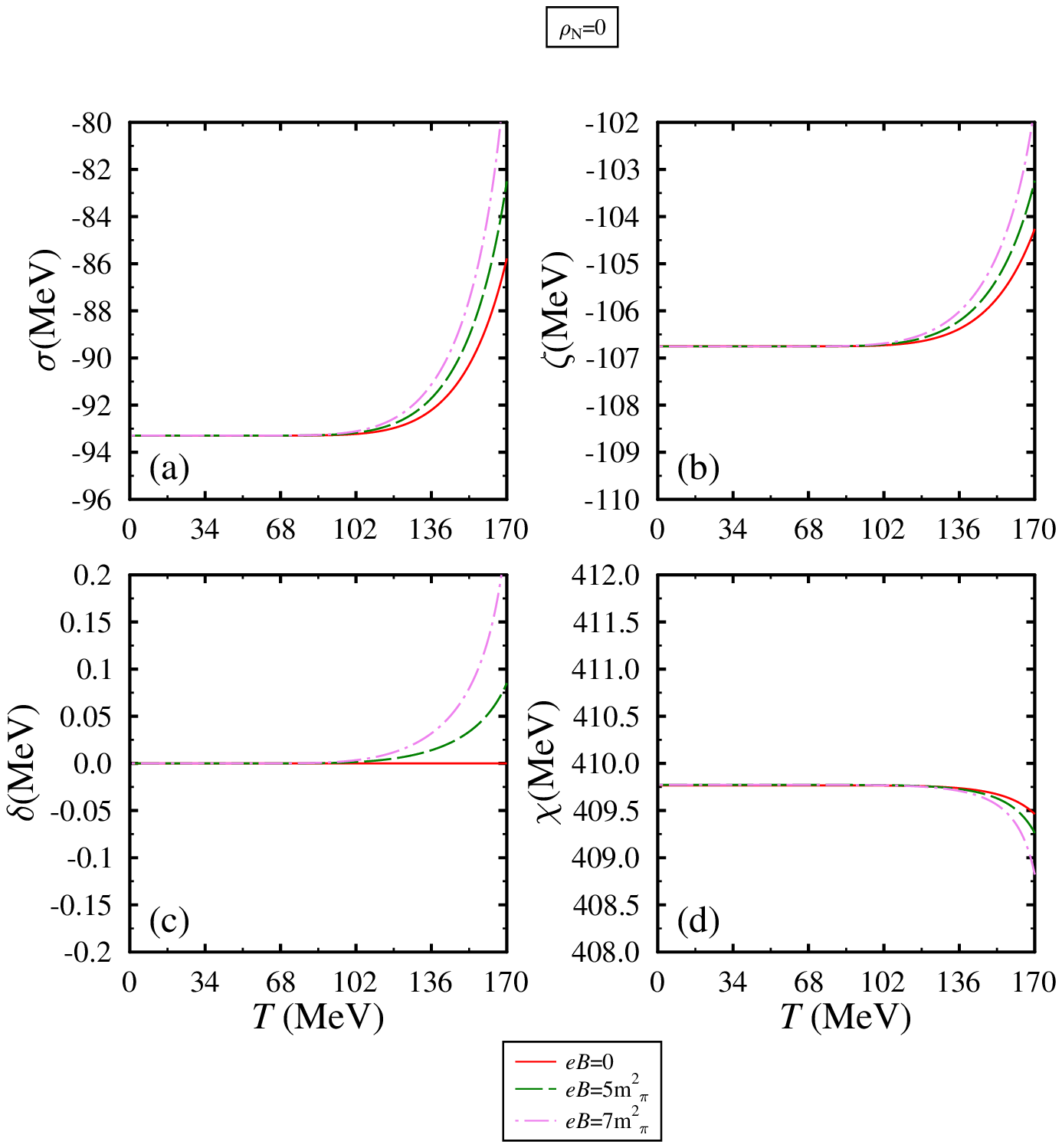}
\caption{(Color online) The scalar fields $\sigma$, $\zeta$, $\delta$ and $\chi$ plotted as a function of temperature $T$, for different values of magnetic field, $B$ at nucleon density $\rho_N$=0.}
\label{ffields0}
\end{figure} 
 
\begin{figure}
\includegraphics[width=16cm,height=21cm]{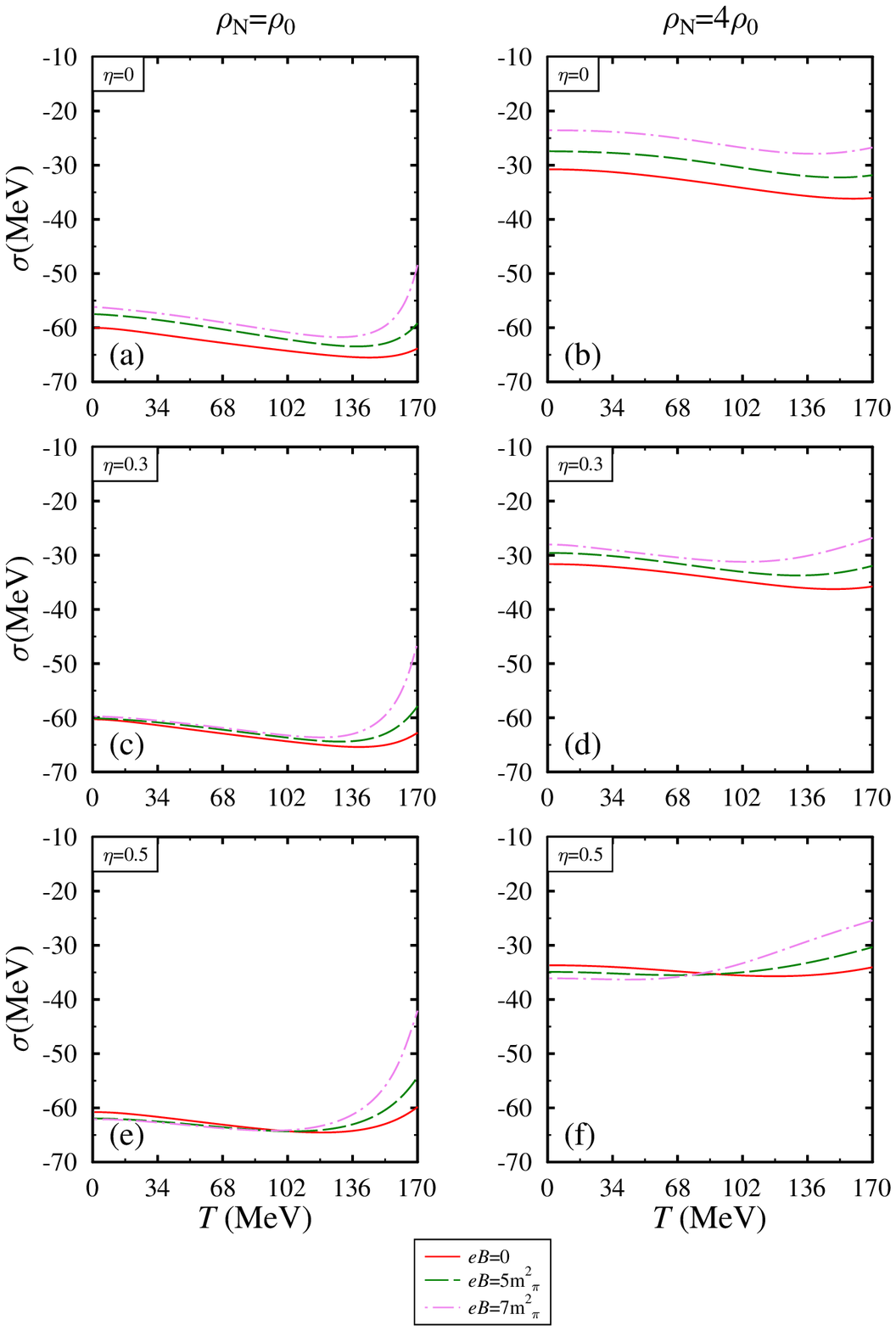}
\caption{(Color online) The scalar field $\sigma$ plotted as a function of temperature $T$, for different values of magnetic field, $B$, nucleon density $\rho_N$ and isospin asymmetry parameter, $\eta$.}
\label{fsigma}
\end{figure}

\begin{figure}
\includegraphics[width=16cm,height=21cm]{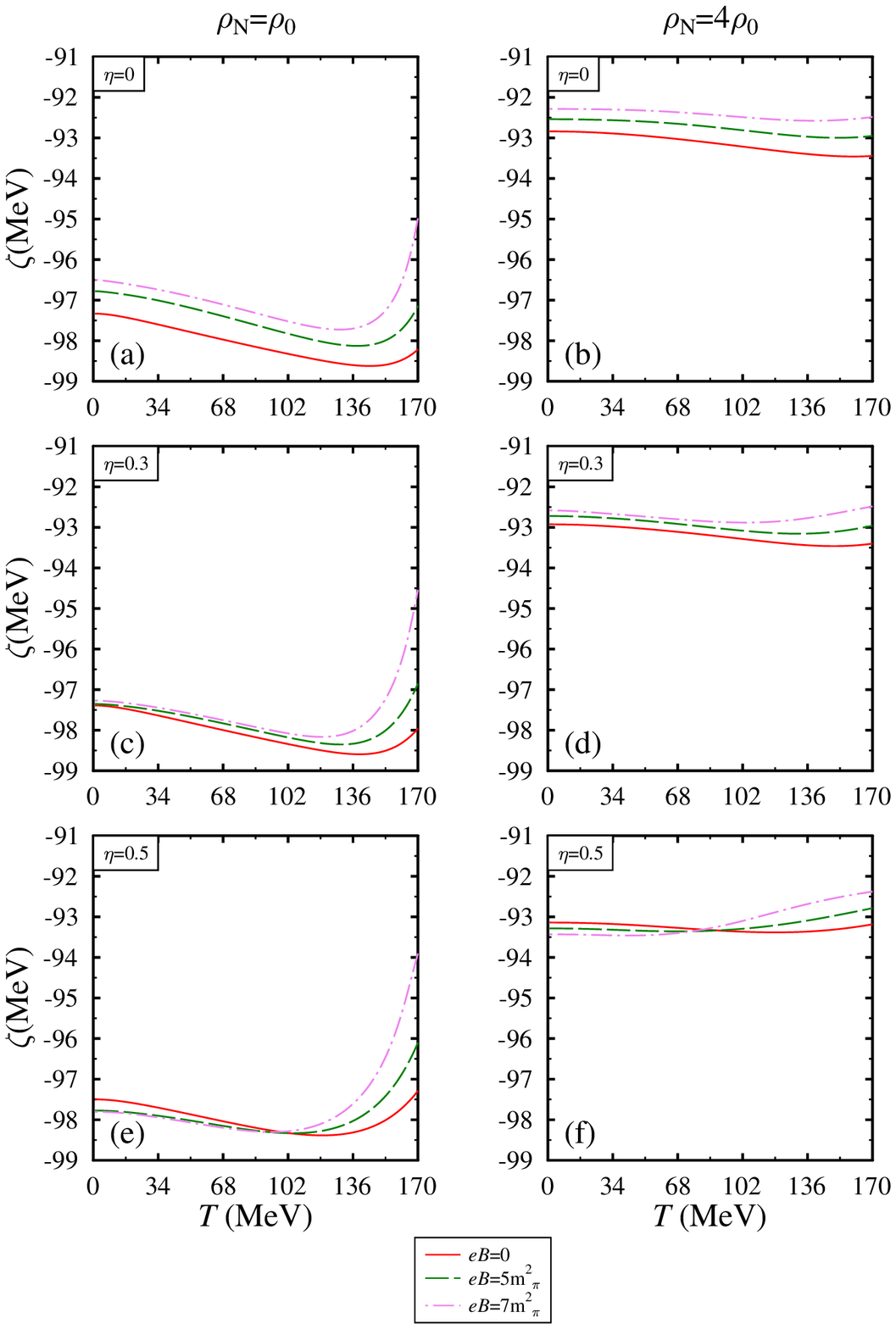}
\caption{(Color online) The scalar field $\zeta$ plotted as a function of temperature $T$, for different values of magnetic field, $B$, nucleon density $\rho_N$ and isospin asymmetry parameter, $\eta$.}
\label{fzeta}
\end{figure}

\begin{figure}
\includegraphics[width=16cm,height=21cm]{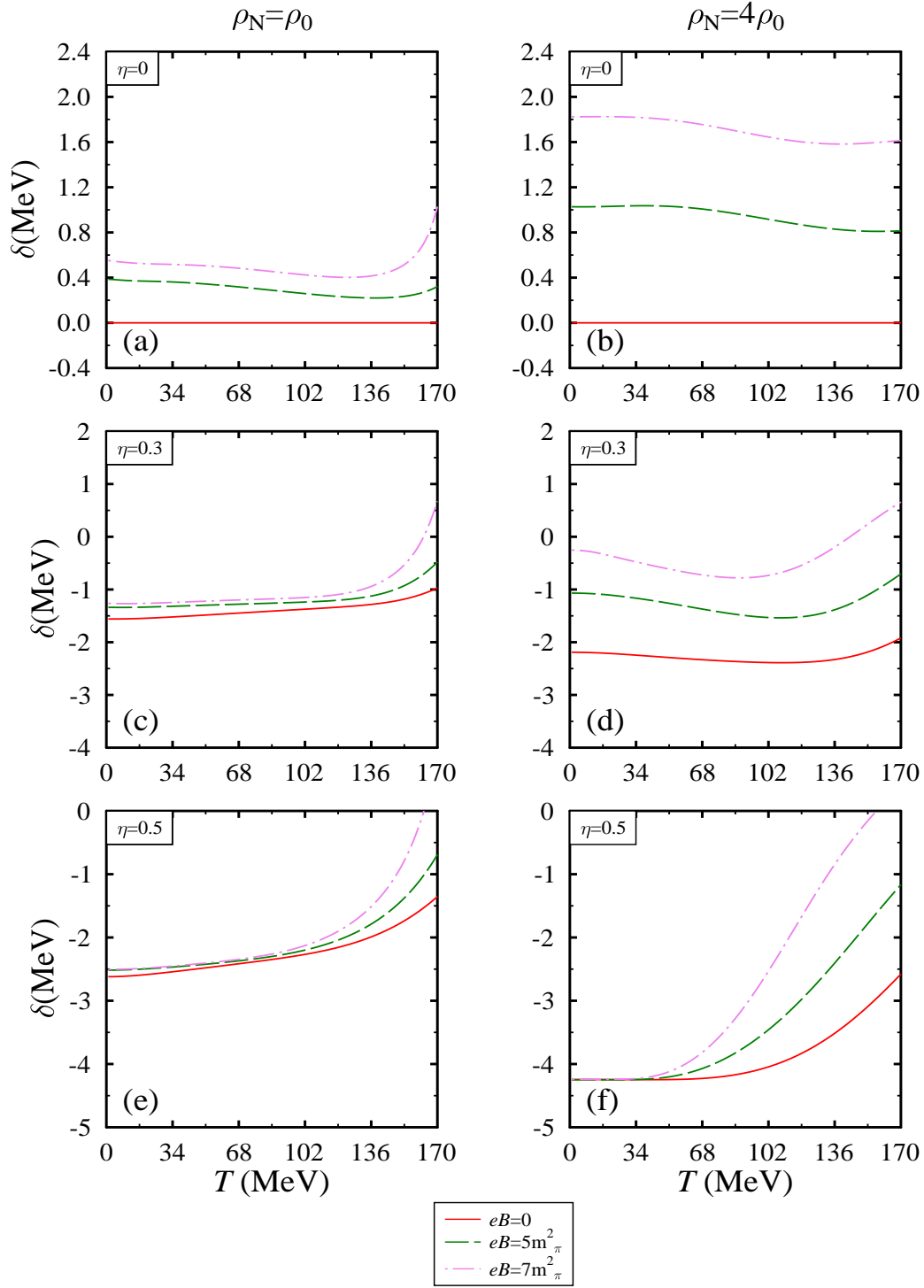}
\caption{(Color online) The scalar field $\delta$ plotted as a function of temperature $T$, for different values of magnetic field, $B$, nucleon density $\rho_N$ and isospin asymmetry parameter, $\eta$.}
\label{fdelta}
\end{figure}

\begin{figure}
\includegraphics[width=16cm,height=21cm]{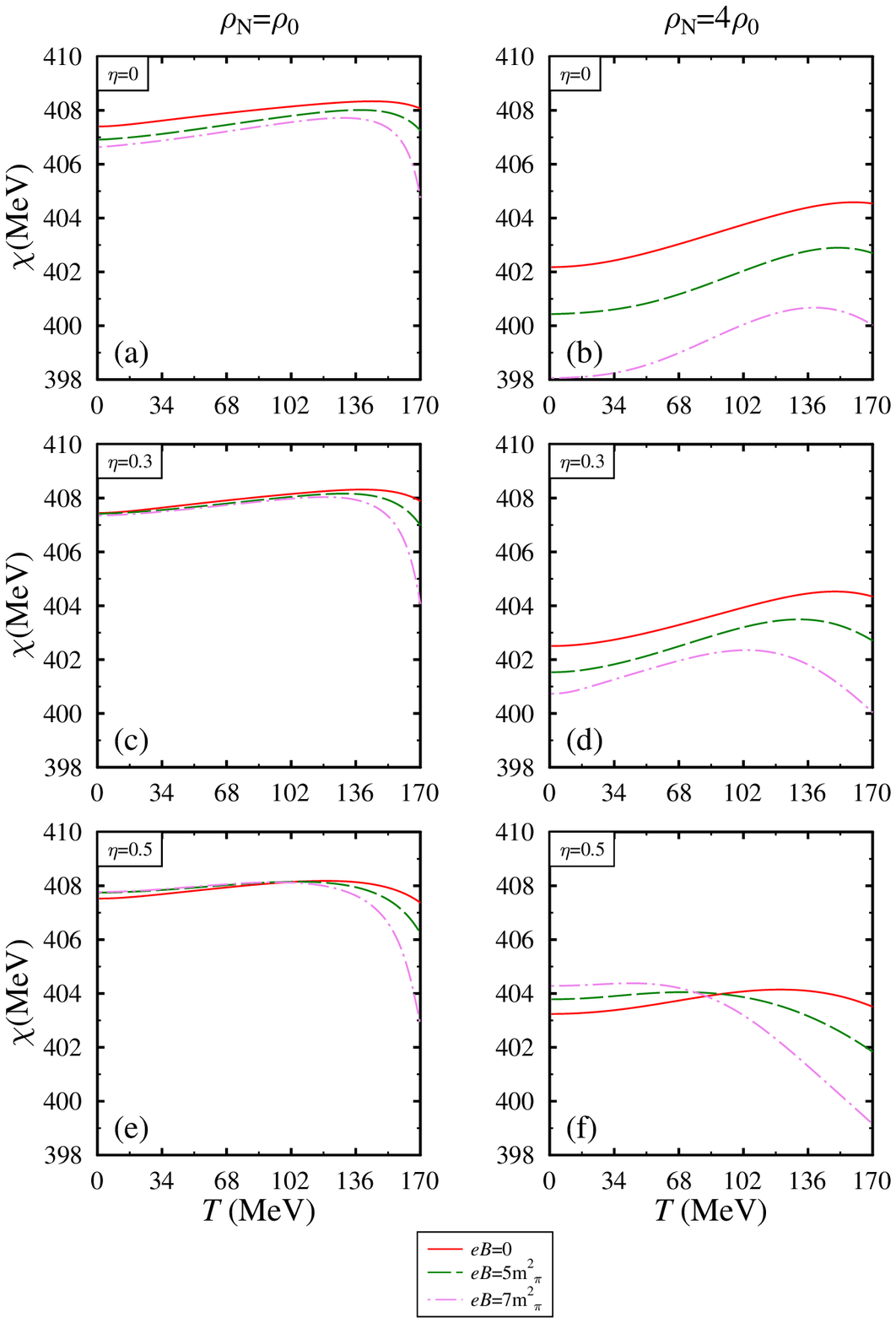}
\caption{(Color online) The dilaton field $\chi$ plotted as a function of temperature $T$, for different values of magnetic field, $B$, nucleon density $\rho_N$  and isospin asymmetry parameter, $\eta$.}
\label{fchi}
\end{figure}

 In Fig. \ref{ffields0}, we show the variation of scalar fields $\sigma,\zeta,\delta$, and $\chi$ as a function of  temperature $T$ at magnetic field B=0, 5${{m^2_{\pi}}}$ and 7${{m^2_{\pi}}}$ (1${{m^2_{\pi}}}$=$ 2.818\times 10^{18}$ gauss = $ 5.48\times 10^{-2}$ GeV$^2$), and nucleon density, $\rho_N$=0. As can be seen from Fig. \ref{ffields0}, at $\rho_N$=0, the effects of magnetic field are more appreciable above certain high value of temperature. For given temperature, the magnitude of scalar fields $\sigma$ and $\zeta$ and the dilaton fields $\chi$ are observed to decrease with increase in magnetic field. Further, in Figs. \ref{fsigma} to \ref{fchi}, we plot these scalar fields  individually as a function of  temperature $T$ at different magnetic field  strength, $eB$ and for finite nucleon density. We show the results at  nucleon densities $\rho_N=\rho_0$ (left column) and 4$\rho_0$ (right column) and isospin asymmetry, $\eta$=0, 0.3 and 0.5.  Here, e and $\rho_0$ are the electrostatic unit of charge and nuclear saturation density respectively.

In Figs. \ref{fsigma} and \ref{fzeta}, we observe that at zero and non-zero values of magnetic field, for finite value of nucleon density $\rho_N$, the magnitude of scalar fields $\sigma$ and $\zeta$ increases with increase in temperature and then start decreasing  after certain value. For example, in symmetric nuclear matter, at $\rho_N$=$\rho_0$ and $B$=0, the magnitudes of both $\sigma$ and $\zeta$ fields increase upto temperature $T$=145 MeV and then start decreasing with further increase in temperature. For finite value of magnetic field $B$, the magnitude of both scalar fields decreases as compared to $B$=0, but follows the same functional dependence with respect to temperature. In symmetric nuclear medium, at nucleon density $\rho_N$=$\rho_0$ and temperature $T$=0 (100) MeV, the values of $\sigma$ field are observed to be -60.02 (-64.19), -57.51 (-62.07) and -56.19 (-60.76) MeV for $eB$=0, 5 ${{m^2_{\pi}}}$ and 7 ${{m^2_{\pi}}}$,  respectively, and  the values of $\zeta$ field are observed as -97.33 (-98.30), -96.78 (-97.80) and -96.50 (-97.50) MeV. At $\rho_N$=4$\rho_0$, and temperature $T$=0 (100) MeV, the values of $\sigma$ field are observed as -30.76 (-34.10), -27.43 (-30.37) and -23.55 (-26.67) MeV for  $eB$=0, 5 ${{m^2_{\pi}}}$ and 7 ${{m^2_{\pi}}}$, respectively, and the values of $\zeta$ field are observed to be -92.84 (-93.20), -92.54 (-92.80) and -92.29 (-92.48) MeV. The decrease in the magnitude of $\sigma$ and $\zeta$ fields as a function of magnetic field strength will result in decrease in the mass of nucleons (see Eq.(\ref{mneff})). Thus, finite magnetic field strength at high temperature will further support the restoration of chiral symmetry and may result in decrease in critical temperature.  Moreover, with the rise of temperature, the value of thermal distribution function given in Eqs.(\ref{dfp}) and (\ref{dfn}) decreases, which results in decrease in the scalar density given by Eqs.(\ref{rhosp}) and (\ref{rhosn}). In addition, at ${\mu^*_i}\neq0$, $i. e.$ for finite density, there are contributions from higher momenta. Thus, the variation of $\sigma$ and $\zeta$ field with respect to temperature reflects the competing effects between the thermal distribution functions and the contributions from higher momenta states \cite{Mishra2004}.  

Comparing the behaviour of scalar fields at finite asymmetry of the medium with the symmetric situation, we observe that isospin asymmetric effect on scalar field $\sigma$ are more appreciable than $\zeta$ field. This is related to the fact that the $\sigma$ field contains non zero isospin quark content, while $\zeta$ field quark content is independent of isospin. Therefore, former is more sensitive to non-zero isospin asymmetry parameter, $\eta$, than the latter ($\zeta$ has little dependence as it is solved along with other scalar fields). For given density and temperature, the decrease in magnitude of scalar fields is observed to be small at $\eta$=0.3 as compared to $\eta$=0, as we increase the external magnetic field strength from zero to finite value. However, as we move to $\eta$=0.5, below T$\cong$80   MeV, the magnitude of scalar field $\sigma$ increases with increase in the magnetic field strength, whereas, above this temperature, the trend is opposite. The reason behind this crossover behaviour is the fact that in pure neutron matter, $\eta$=0.5, at a particular value of temperature, neutron scalar density, $\rho^{s}_n$ decrease more slowly as compared to proton scalar density, $\rho^{s}_p$. The net effect of $\rho^{s}_n$ and $\rho^{s}_p$ results in the above behavior of $\sigma$ and $\zeta$ fields.
In the present investigation, the variation of scalar fields and nucleon mass at finite density, temperature and external magnetic field are in accordance with the  calculations done under non-linear Walecka model \cite{Rabhi2011}. In this article, authors have studied the properties of stellar matter at finite density, tempearature and uniform magnetic field.

The scalar isovector field  $\delta$ having contribution due to isospin asymmetry of the medium is plotted in Fig.  \ref{fdelta}. It is observed that at $\eta$=0, the $\delta$ field is zero for magnetic field $B$=0, but non-zero for finite magnetic field strength. This is because the $\delta$ field is determined by the difference between the scalar densities of neutron and proton. For zero magnetic field, the value of $\delta$ field is zero as scalar density of proton is equal to scalar density of neutron (see Eqs. (\ref{rhov0}) and (\ref{rhos0})). On the other hand, in the presence of magnetic field, due to charged nature of proton, Landau quantization occurs, this leads to the inequality $\rho^{s}_p\neq\rho^{s}_n$ (see Eqs. (\ref{rhovp}) and (\ref{rhosn})), hence non-zero value of the $\delta$ is obtained. In Fig. \ref{fchi}, the dilaton field $\chi$  is plotted, which is introduced in the chiral $SU(3)$ model to preserve broken scale invariance property of QCD at tree level \cite{Schechter1980}. It is observed that the magnitude of dilaton field $\chi$ shows behaviour similar to $\sigma$ and $\zeta$ fields. It first increases up to certain temperature $T$, then decreases with further increase in temperature. It is observed from the figure that the dilaton field $\chi$ varies least as compared to other mesonic fields. Therefore, frozen glueball limit is assumed in many research papers on chiral $SU(3)$ model \cite{Reddy2018,Papazoglou1999}. In the lower density region, the effects of magnetic field on $\chi$ field are very less, whereas at high density these become significant.

\subsection{In-medium Gluon Condensate and Twist-2 Gluon Operator} 

\begin{figure}
\includegraphics[width=16cm,height=21cm]{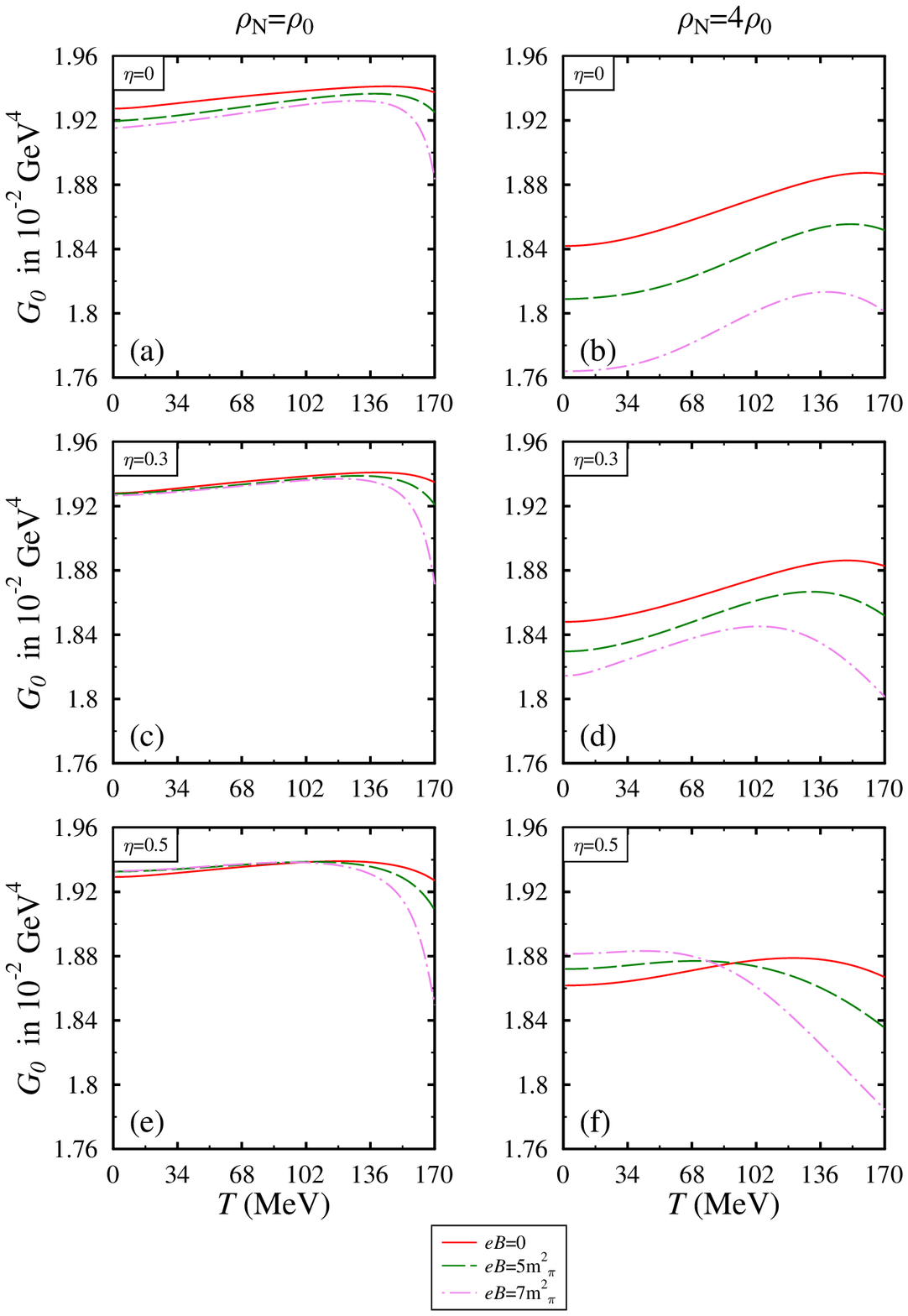}
\caption{(Color online) The  scalar gluon condensate $G_0$  describing the trace part of energy momentum tensor plotted as a function of temperature $T$, for different values of magnetic field, $B$, nucleon density $\rho_N$  and isospin asymmetry parameter, $\eta$.}
\label{fG0}
\end{figure}

\begin{figure}
\includegraphics[width=16cm,height=21cm]{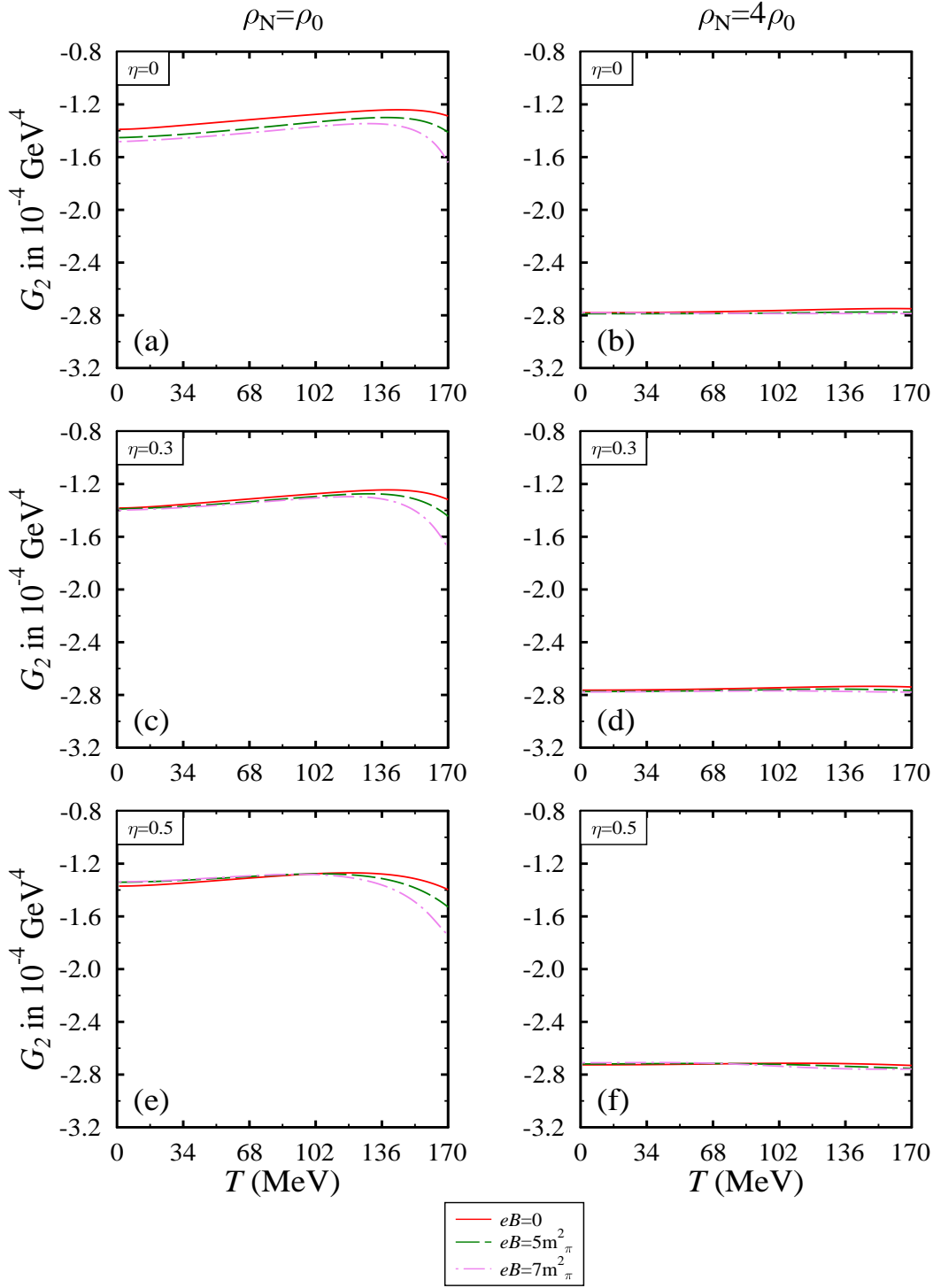}
\caption{(Color online) The twist 2 gluon operator
$G_2$ describing the nontrace part of energy momentum tensor plotted as a function of temperature $T$, for different values of magnetic  field, $B$, nucleon density $\rho_N$ and isospin asymmetry parameter, $\eta$.}
\label{fG2}
\end{figure}

\begin{figure}
\includegraphics[width=16cm,height=21cm]{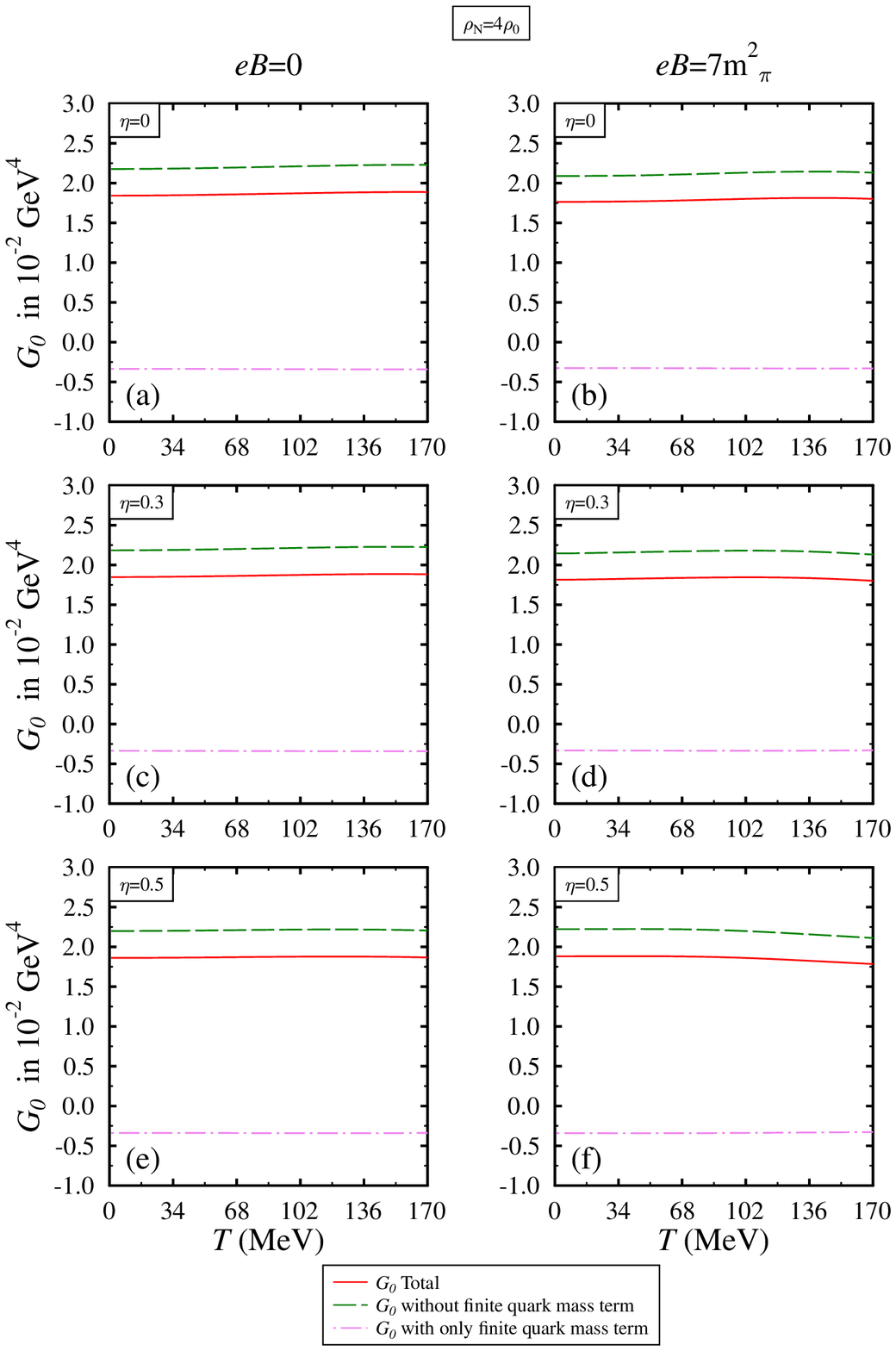}
\caption{(Color online) The individual terms of scalar gluon condensate $G_0$  plotted as a function of temperature $T$, for different values of magnetic field, $B$, and isospin asymmetry parameter, $\eta$ at nucleon density $\rho_N$=$4\rho_0$.}
\label{fG0t}
\end{figure}

\begin{figure}
\includegraphics[width=16cm,height=21cm]{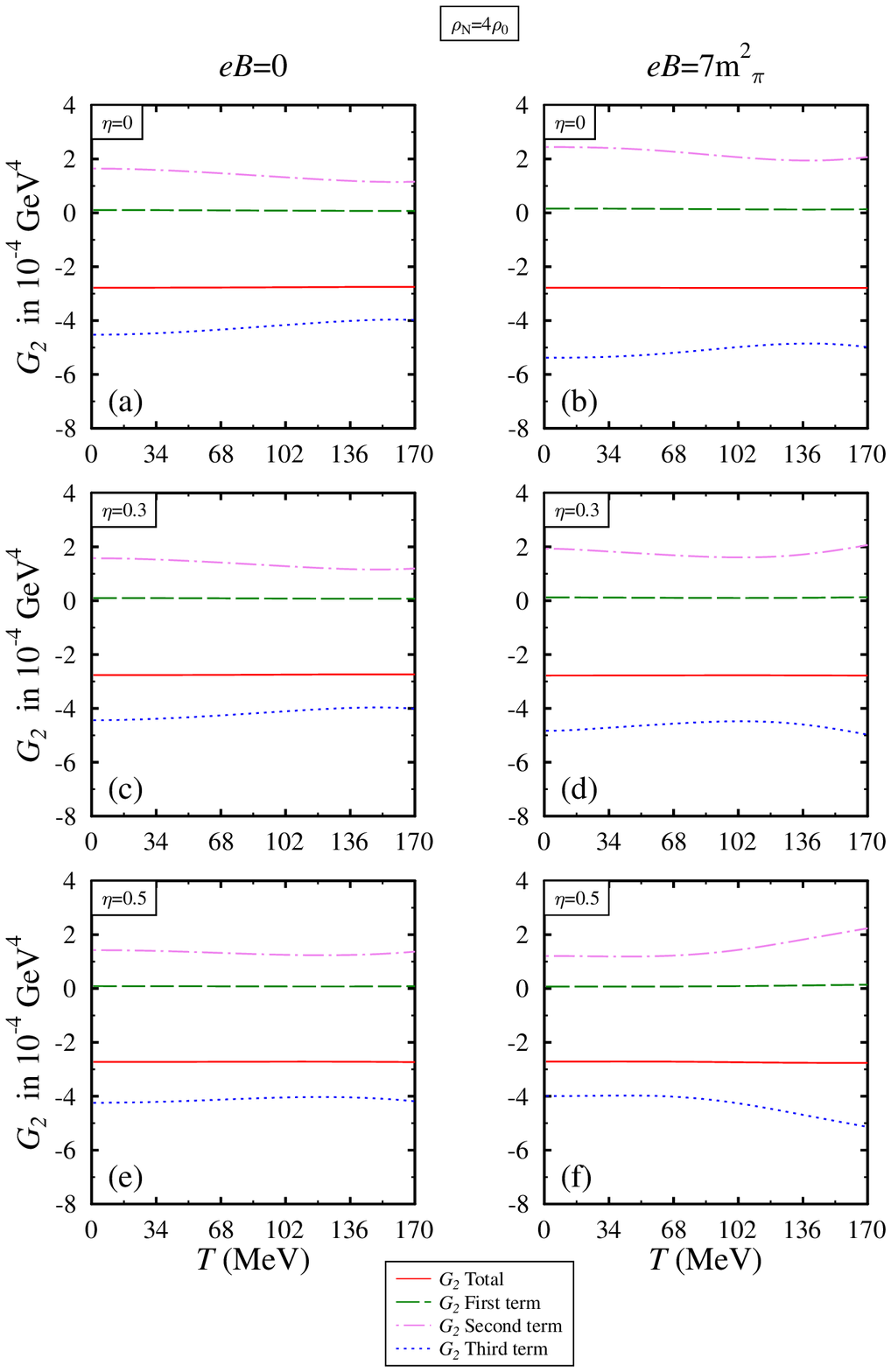}
\caption{(Color online) The individual terms of tensorial gluon operator $G_2$  plotted as a function of temperature $T$, for different values of magnetic field, $B$, and isospin asymmetry parameter, $\eta$ at nucleon density $\rho_N$=$4\rho_0$.}
\label{fG2t}
\end{figure}

In Figs. \ref{fG0} and \ref{fG2}, we show the variations of scalar gluon condensate $G_0$ (Eq.(\ref{chiglum}))  as well as  the twist-2 gluon operator $G_2$ (Eq.(\ref{g2approx}))  with temperature for different values of magnetic field, nucleon density and isospin asymmetry parameter. The scalar gluon condensate, $G_0$ consists of two terms. The first term has the dependence on the fourth power of $\chi$ field, whereas the second term contributes the finite quark mass effects from the energy-momentum tensor, which have the $\sigma$ and $\zeta$ field dependence. On the other hand,  the twist-2 gluon operator, $G_2$ has contribution of three terms, which further have the dependency on scalar fields. Although, the condensates $G_0$  and $G_2$ depends upon  all scalar fields (see Eqs.(\ref{chiglum}) and (\ref{g2approx})), but the   observed behaviour  resembles $\chi$ field due to its fourth power dependence.   For better understanding of condensate's behaviour, the different terms of $G_0$  and $G_2$ are plotted in Figs. \ref{fG0t} and \ref{fG2t} respectively. Considering finite quark mass term contribution to scalar gluon condensates, in symmetric nuclear medium, at temperature $T$=0 (100) MeV, the values of $G_0$, in units of $10^{-2}$ GeV${^4}$, are observed to be 1.842 (1.871), 1.809 (1.838) and 1.764 (1.800) at $eB$=0, 5${{m^2_{\pi}}}$ and 7${{m^2_{\pi}}}$, respectively. However, in massless quark limit (Eq.(\ref{chiglu})) these values changes to 2.177 (2.210), 2.139 (2.172) and 2.089 (2.130).  In addition, the tensorial gluon operator, $G_2$ varies very less in high density as compared with low density as a function of temperature. This behaviour reflects the competing effects between the second and third term of Eq.(\ref{g2approx}) plotted in Fig. \ref{fG2t}. The crossover behaviour of condensates in highly asymmetric nuclear medium reflects the $\chi$ field, which is solved along with the $\sigma$, $\zeta$ and $\delta$ fields. 
For example, in symmetric nuclear matter, at $\rho_N$=4$\rho_0$ and temperature $T$=0 (100) MeV, the values of $\chi$ field are observed to be 402.2 (403.5), 400.4 (402) and 398.1 (400) MeV for $eB$=0, 5${{m^2_{\pi}}}$ and 7${{m^2_{\pi}}}$ respectively and for $\eta=0.5$, these values changes to 403.2 (404.1), 403.8 (403.9) and 404.3 (403.3). In vacuum, the twist-2 gluon operator vanishes due to the Lorentz symmetry. But in the current investigation, the presence of finite density, temperature and magnetic field breaks the Lorentz symmetry, hence  non-vanishing value of $G_2$ is obtained \cite{Morita2008,Klingl1999}. The observed behaviour of $G_0$ and $G_2$  as a function of temperature at zero magnetic field and nucleon density is consistent with the condensates behaviour observed in Ref. \cite{Morita2008}.

\subsection{In-medium Masses of $J/\psi$ and $\eta_c$ } 

\begin{figure}
\hspace{-2cm} 
\includegraphics[width=14cm,height=14cm]{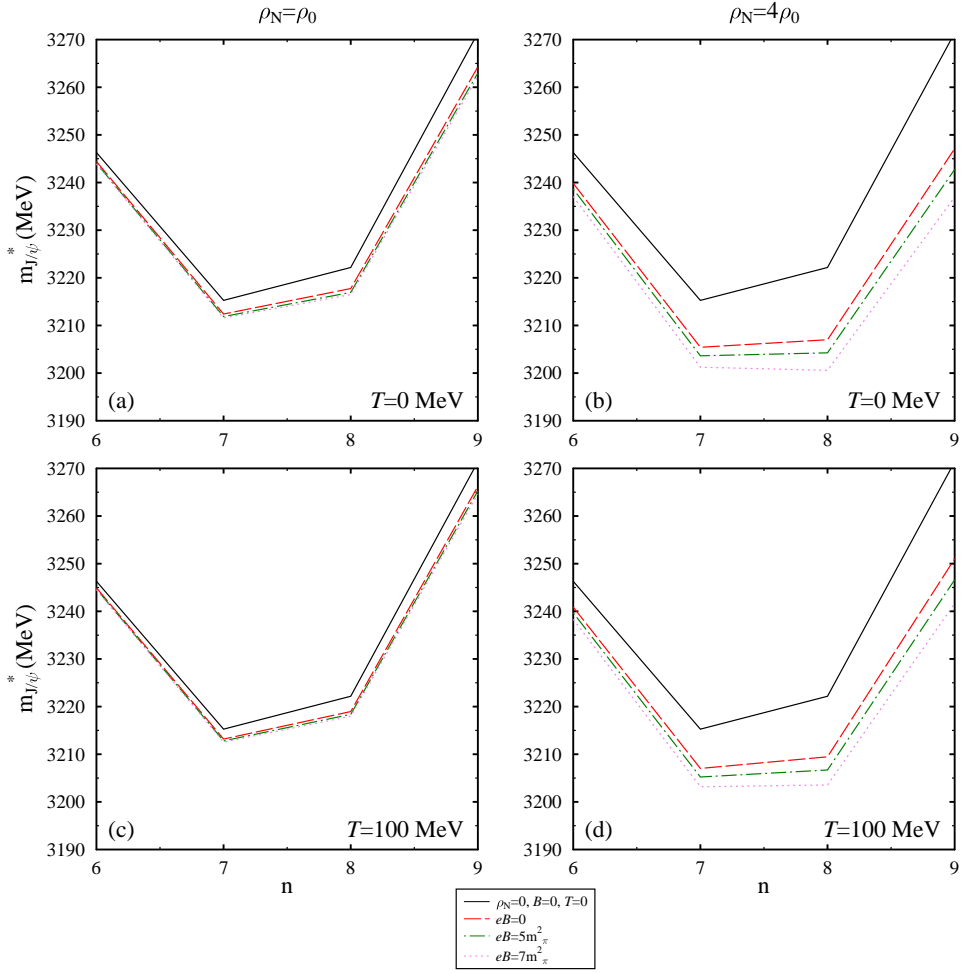}
\caption{(Color online)The in-medium mass of $J/\psi$ meson plotted as a function of n for symmetric nuclear matter at different values of  temperature, magnetic field and density.}
\label{jpsieta0}
\end{figure}

\begin{figure}[htbp]
\hspace{-2cm} 
\includegraphics[width=14cm,height=14cm]{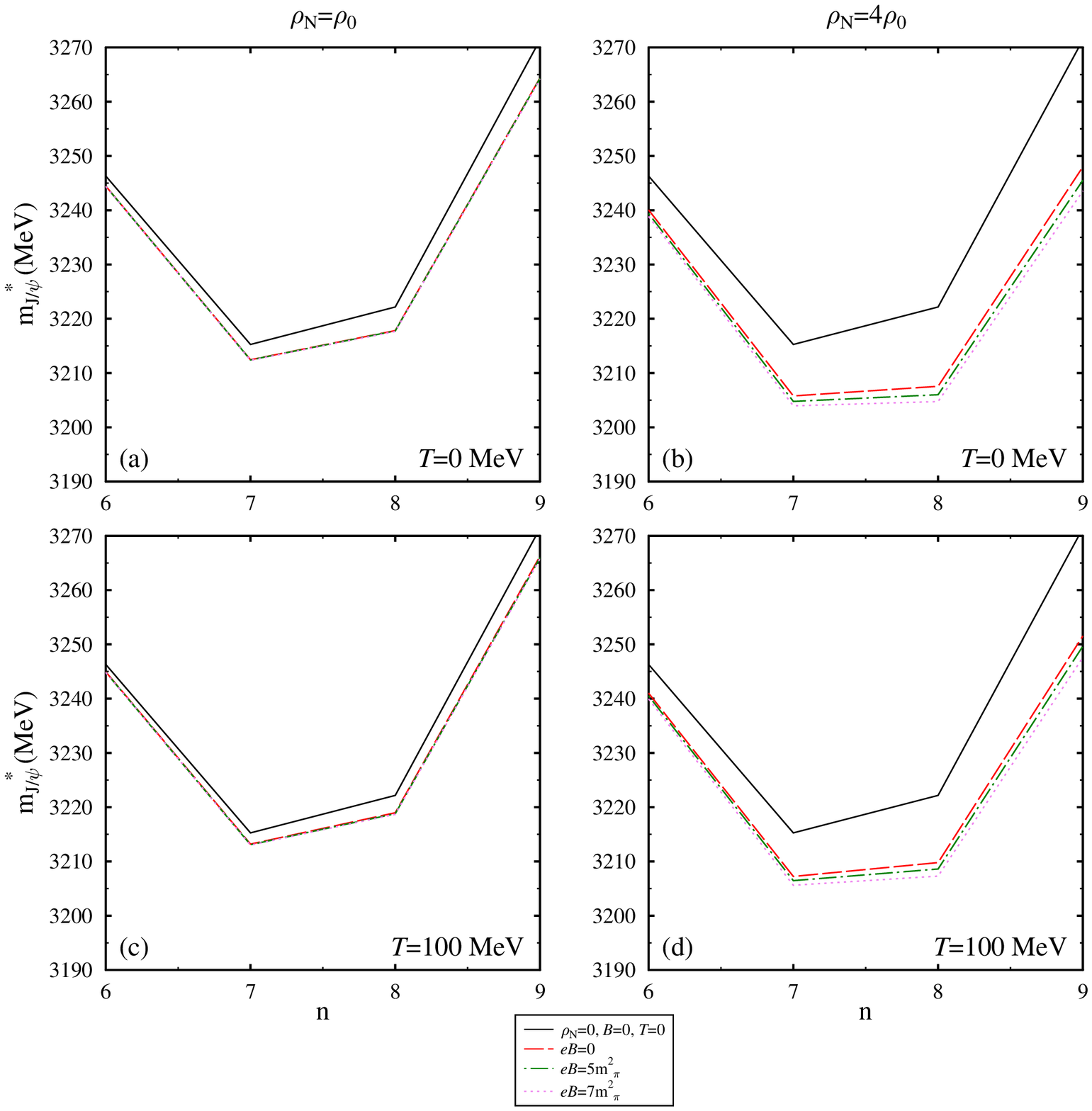}
\caption{(Color online)The in-medium mass of $J/\psi$ meson plotted as a function of n for asymmetric nuclear matter, $\eta$=0.3  at different values of temperature, magnetic field and density.}
\label{jpsieta3}
\end{figure}

\begin{figure}
\hspace{-2cm} 
\includegraphics[width=14cm,height=14cm]{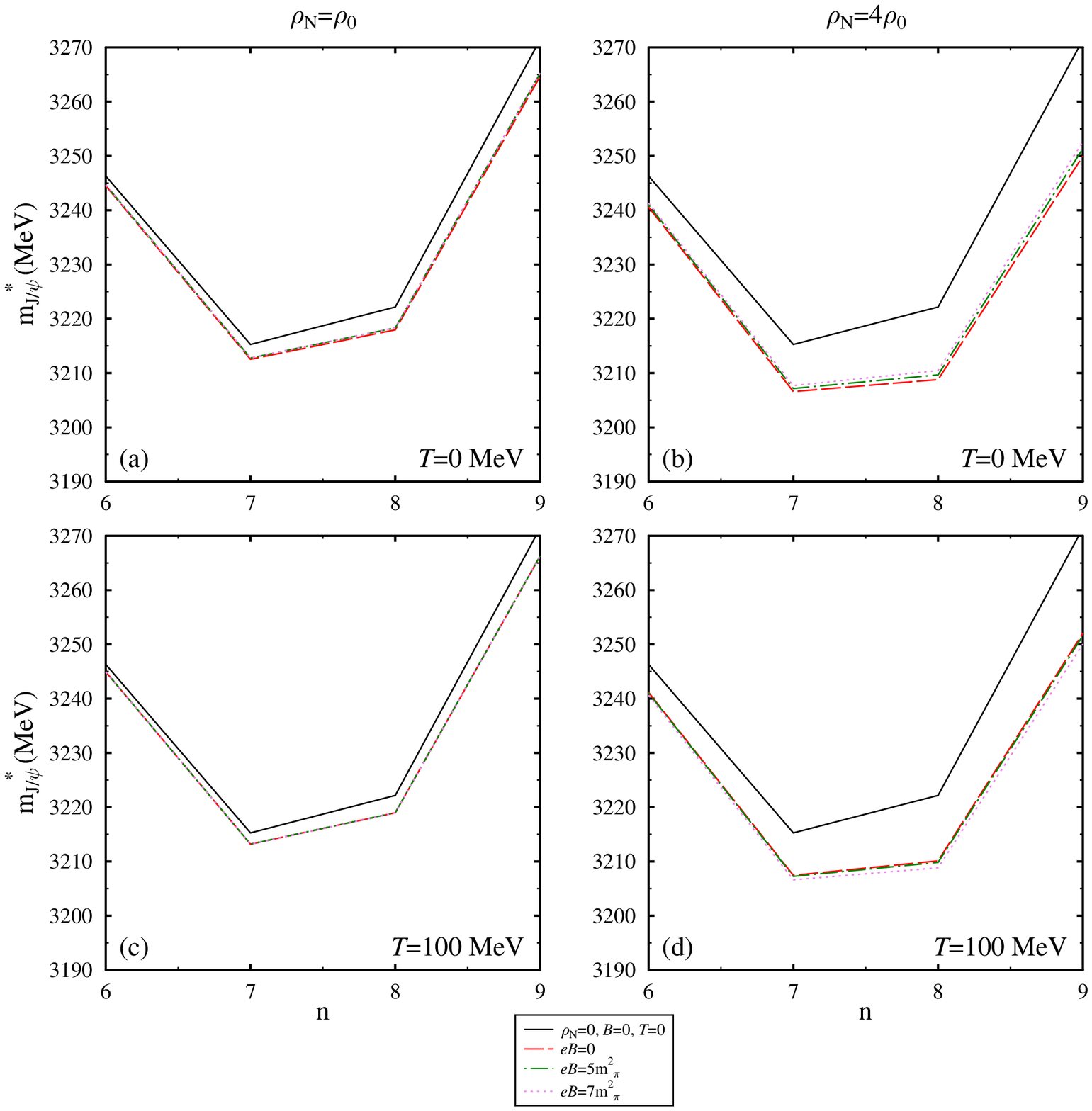}
\caption{(Color online)The in-medium mass of $J/\psi$ meson plotted as a function of n for asymmetric nuclear matter, $\eta$=0.5  at different values of temperature, magnetic field and density.}
\label{jpsieta5}
\end{figure}

\begin{figure}
\hspace{-2cm} 
\includegraphics[width=14cm,height=14cm]{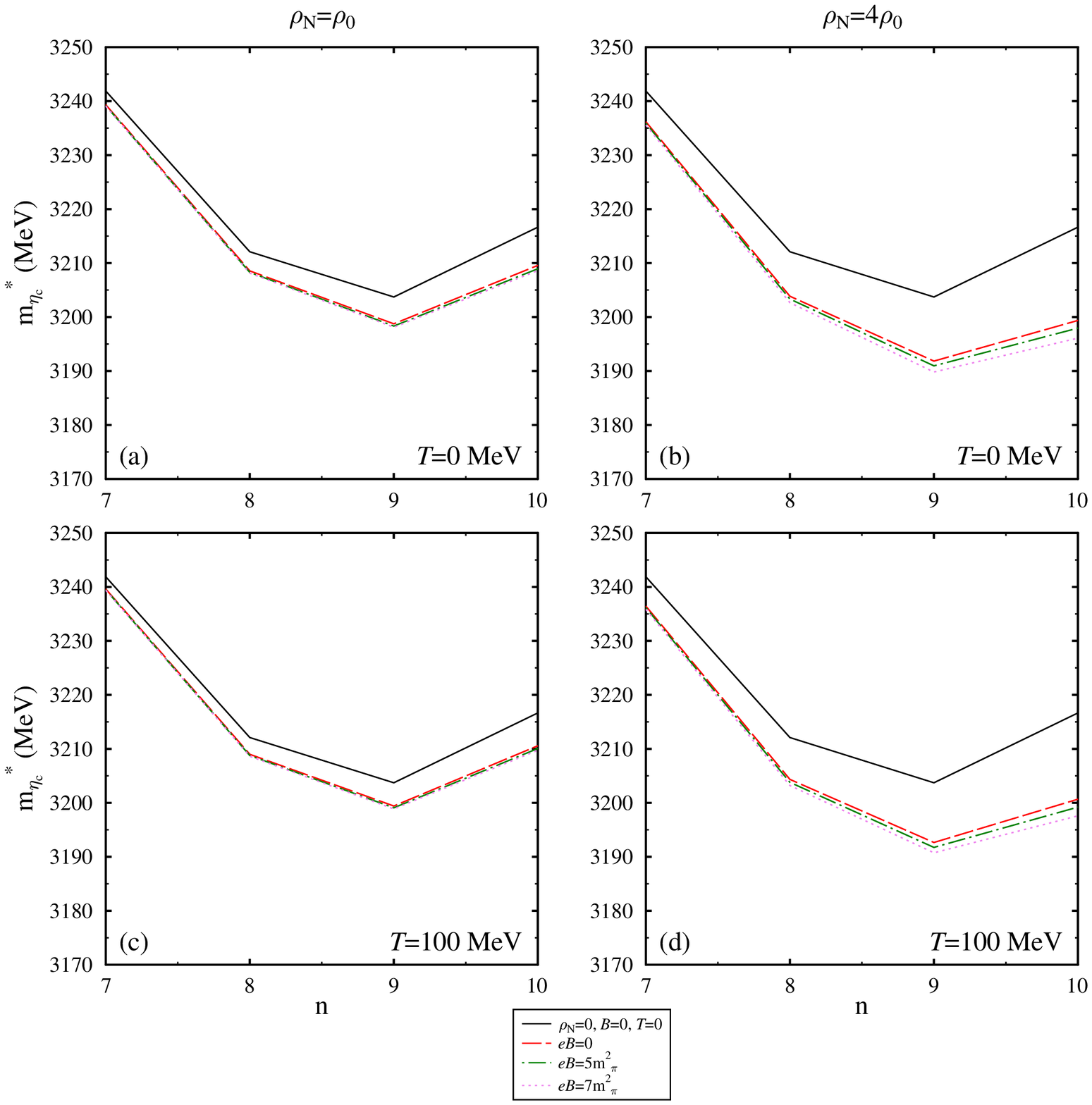}
\caption{(Color online)The in-medium mass of $\eta_c$ meson plotted as a function of n for symmetric nuclear matter  at different values of temperature, magnetic field and density.}
\label{etaeta0}
\end{figure}

\begin{figure}
\hspace{-2cm} 
\includegraphics[width=14cm,height=14cm]{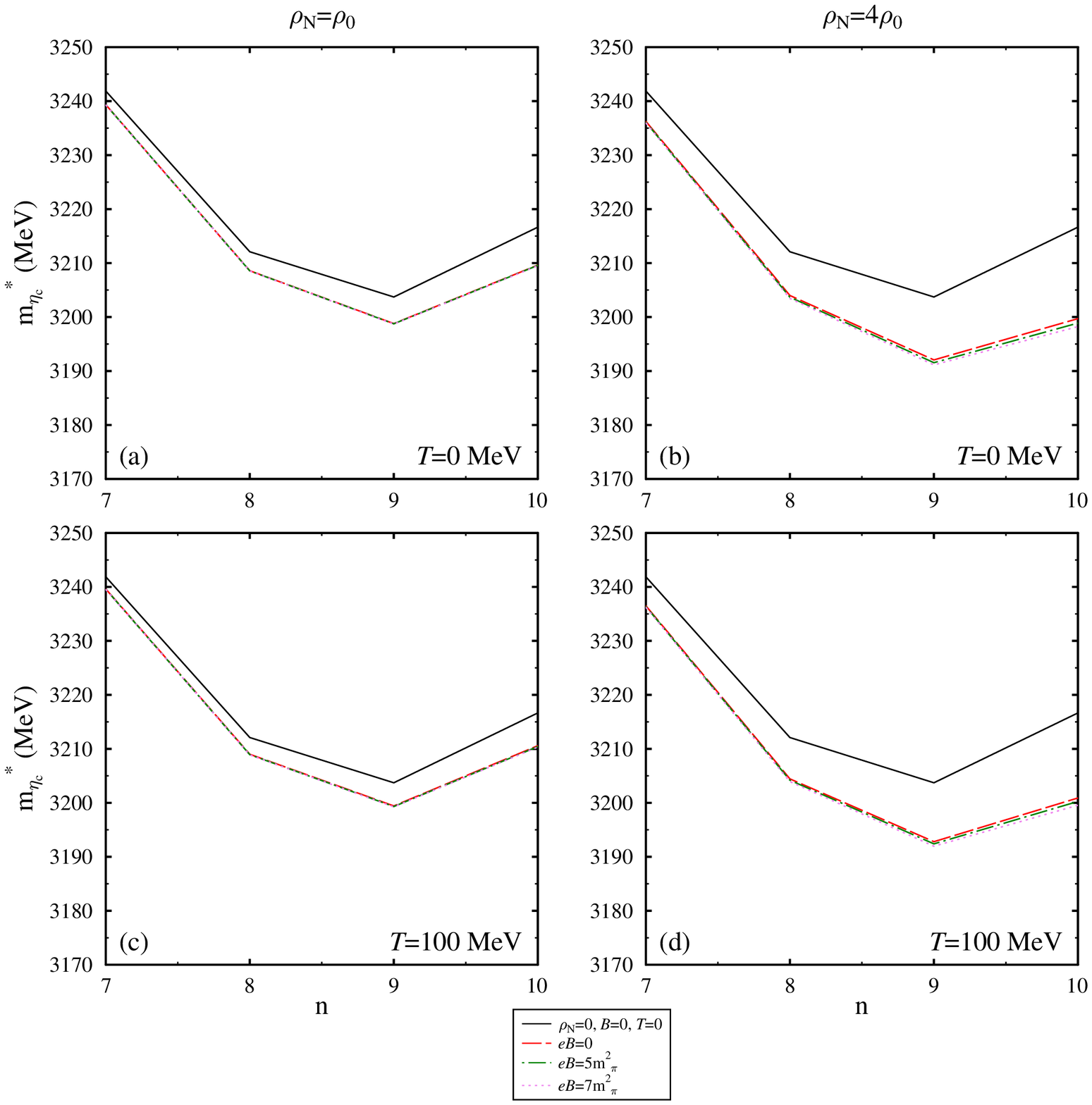}
\caption{(Color online)The in-medium mass of $\eta_c$ meson plotted as a function of n for asymmetric nuclear matter, $\eta$=0.3  at different values of temperature, magnetic field and density.}
\label{etaeta3}
\end{figure}

\begin{figure}
\hspace{-2cm} 
\includegraphics[width=14cm,height=14cm]{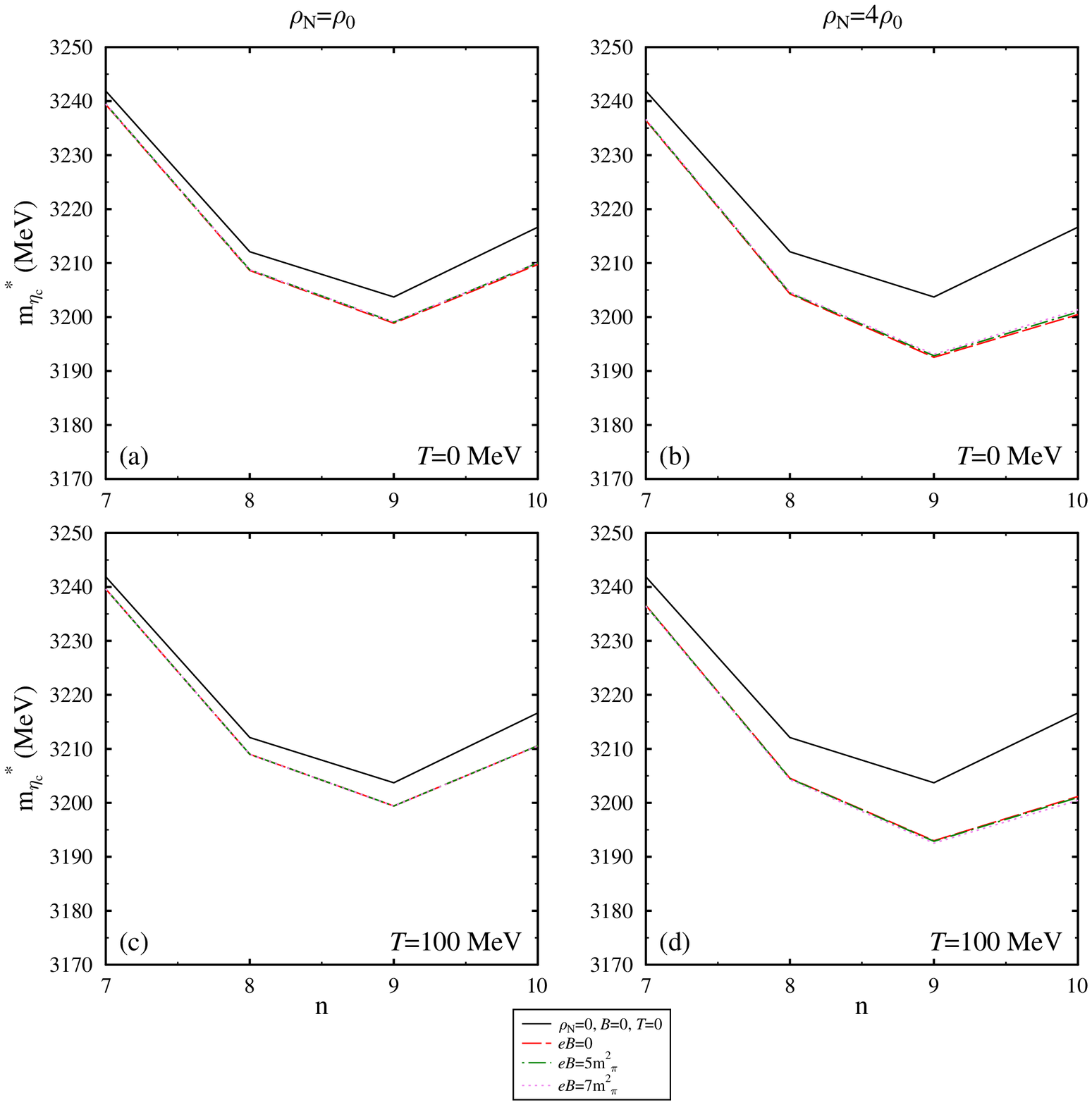}
\caption{(Color online)The in-medium mass of $\eta_c$ meson plotted as a function of n for asymmetric nuclear matter, $\eta$=0.5  at different values of temperature, magnetic field and density.}
\label{etaeta5}
\end{figure}

In this section, we calculate the in-medium mass shift of $J/\psi$  and $\eta_c$ mesons, using medium modified scalar and tensorial gluon condensate, by applying QCD sum rules. In this application, we use the moment in the range $5\leq n\leq 9$ and $7\leq n \leq 10$ for $J/\psi$  and $\eta_c$ mesons respectively to get the mass in stable region. We have used $\xi$=1 and corresponding running charm quark mass, $m_c$=1228.52 MeV as well as parameter, $\alpha_s$=0.2636. In Eqs.(\ref{phibglu}) and (\ref{phicglu}), we obtain the value of $\phi_b$ and $\phi_c$ using medium modified gluon condensates within the chiral $SU(3)$ model and are used to calculate moment $M_{n}^{J}$, and hence the in-medium charmonium mass ${m^*_C}$, under QCD sum rules. In cold symmetric nuclear matter,  at $\rho_N$=$\rho_0(4\rho_0)$ the values of $\phi_b$ in multiple of $10^{-3}$, are found to be 2.30 (2.20), 2.29 (2.18) and 2.29 (2.14) at  $eB$=0, 5${{m^2_{\pi}}}$ and 7${{m^2_{\pi}}}$ respectively. These values can be compared with the values of $\phi_b$ equal to $1.7\times10^{-3}$ and $1.6\times10^{-3}$ at vacuum and nuclear saturation density in Ref. \cite{Klingl1999} . In the present investigation,  for the same conditions of medium, but $T$=100 MeV, above values of $\phi_b$ changes to 2.29 (2.19), 2.28 (2.15) and 2.27 (2.09). The values of $\phi_c$, in multiple of $10^{-5}$, at  $eB$=0, 5${{m^2_{\pi}}}$ and 7${{m^2_{\pi}}}$ are observed to be -5.01 (-1.003), -5.24 (-1.006) and -5.35 (-1.009) at $\rho_N$=$\rho_0$ and $T$=0.  We plot in-medium mass of $J/\psi$ in Figs. \ref{jpsieta0}, \ref{jpsieta3}, and \ref{jpsieta5}  for isospin asymmetry $\eta$=0, 0.3 and 0.5 respectively. In-medium mass of $\eta_c$ is  plotted in Figs. \ref{etaeta0}, \ref{etaeta3}, and \ref{etaeta5} for same isospin parameters. We show the results for  $J/\psi$ and $\eta_c$  at $T$=0,100 MeV and $\rho_N$=$\rho_0$,$4\rho_0$ and for magnetic field $eB$=0, 5${{m^2_{\pi}}}$ and 7${{m^2_{\pi}}}$. In each subplot the results are compared with vacuum situation, $i.e.$ $\rho_N$=$B$=$T$=0.  In table \ref{table_mass-shift}, we summarize the values of mass-shift of $J/\psi$ and $\eta_c$  at $eB$=5${{m^2_{\pi}}}$ and 7${{m^2_{\pi}}}$ from the values at $B$=0.

From Fig. \ref{jpsieta0} we observe that, for given density and temperature, the finite magnetic field causes more decrease in masses of $J/\psi$ and $\eta_c$ mesons.  However, we observe less decrease in mass of $J/\psi$ and $\eta_c$, as we move from $T$=0 to $T$=100 MeV. The negative mass-shift due to magnetic field is more in  high density regime. For example, at $\rho_N$=$\rho_0$, $\eta$=0 and $T$=0, the mass-shift of $J/\psi$ and $\eta_c$ is -3.39 (-3.68) and -4.63 (-5.07) MeV for $eB$=${5{m^2_{\pi}}}$ ($7{{m^2_{\pi}}}$) with respect to vacuum mass respectively, whereas at 4$\rho_0$, the mass-shift modifies to -11.63 (-14.68) and -11.97 (-12.96) MeV . This is due to the fact that at high density the values of gluon condensates decrease more with the increase of magnetic field, which further modify $\phi_b$ and $\phi_c$ and finally $J/\psi$ and $\eta_c$.  In Ref. \cite{Jahan2018}, using chiral $SU(3)$ model with QCD second order stark effect, authors have reported the $J/\psi$ mass-shift in finite magnetic field with respect to vacuum at zero temperature. In this article, it was observed that, at $\eta$=0, and $\rho_N$=$\rho_0$, the mass-shift is -8.16 MeV for both $eB$=4${{m^2_{\pi}}}$ and 8${{m^2_{\pi}}}$ with respect to vacuum mass . In  Ref. \cite{Cho2015}, using QCD sum rules, the mass shift of $J/\psi$ and $\eta_c$  is calculated in strong magnetic field at zero density and temperature and observed to decrease with increase in $eB$.

For given density and temperature, as we increase the isospin asymmetry of the medium from $\eta$=0 to 0.5, less negative mass-shift is observed at finite magnetic field from zero magnetic field situation. In other words, finite isospin asymmetry of the medium causes an increase in the mass of charmonium states. The observed behaviour is again due to such dependence of gluon condensates on the isospin asymmetry of nuclear medium.
  At $\eta$=0.5, the mass-shift can be compared with value, -7.86 MeV at $eB$=${8{m^2_{\pi}}}$ obtained in Ref. \cite{Jahan2018}.  In the present investigation, at  $\eta=0.5$, positive mass-shift is observed in low temperature regime and is given in table \ref{table_mass-shift}. The above variation is due to the observed crossover of scalar densities of nucleons and scalar fields $\sigma$ and $\zeta$ as shown in Fig. \ref{fsigma} and \ref{fzeta}.

In Figs. \ref{jpsieta0} and \ref{etaeta0}, when there is no external  magnetic field and isospin asymmetry, it is observed that the mass-shift of $J/\psi$ and $\eta_c$ mesons with respect to vacuum mass  are -2.86 (-2.08) MeV and -4.96 (-4.30) MeV at $T$=0 (100) MeV and $\rho_N$=$\rho_0$.  These values of mass-shift can also be compared with the mass-shift of $J/\psi$ and $\eta_c$ mesons, having -7 and -5 MeV, respectively, investigated in QCD sum rules under linear density approximation \cite{Klingl1999}.

  In the next section, we will now summarize our present work.

\begin{table}[]
\begin{tabular}{|l|l|l|l|l|l|l|l|l|l|l|l|l|l|}
\hline
\multirow{3}{*}{}  & \multirow{3}{*}{} & \multicolumn{4}{c|}{$\eta=0$}                          & \multicolumn{4}{c|}{$\eta=0.3$}                         & \multicolumn{4}{c|}{$\eta=0.5$}                         \\ \cline{3-14} 
                   &          $\Delta m_B$          & \multicolumn{2}{c|}{$T$=0} & \multicolumn{2}{c|}{$T$=100} & \multicolumn{2}{c|}{$T$=0} & \multicolumn{2}{c|}{$T$=100} & \multicolumn{2}{c|}{$T$=0} & \multicolumn{2}{c|}{$T$=100} \\ \cline{3-14} 
                 &  &$\rho_0$ &$4\rho_0$ &$\rho_0$  &$4\rho_0$ &$\rho_0$ &$4\rho_0$ &$\rho_0$  &$4\rho_0$ &$\rho_0$ &$4\rho_0$ &$\rho_0$ &$4\rho_0$ \\ \hline
\multirow{2}{*}{$J/\psi$} &        $\Delta m_{B_{50}}$            &-0.54 &-1.79 &-0.39 & -1.80 & -0.03 &-1.01 &-0.12 &-0.77&0.24 &0.57 &-0.01 &-0.18 \\  \cline{2-14}
                   &         $\Delta m_{B_{70}}$ &-0.84 & -4.84 &-0.65 &-3.85 &-0.11 &-1.83 & -0.07& -1.63 & 0.27 &1.10 &-0.02& -0.83 \\  \hline
\multirow{2}{*}{$\eta_c$} &     $\Delta m_{B_{50}}$                           &-0.44&-0.88&-0.36&-0.92&-0.04&-0.52&-0.11&-0.41&
0.20&0.29&-0.01&-0.11\\  \cline{2-14}
                   &        $\Delta m_{B_{70}}$                        &-0.68&-2.02&-0.59&-1.92&-0.10&-0.92&-0.18&-0.85
&0.23&0.57&-0.01&-0.46\\ 

\hline
\end{tabular}
\caption{In the above table,  we tabulate the  effect of magnetic field on the mass-shifts of  $J/\psi$ and $\eta_c$ mesons. Here, $\Delta m_{B_{50}}$ represents mass-shift between $eB$=5${{m^2_{\pi}}}$ and B=0 (similar for $\Delta m_{B_{70}}$ ). In addition, temperature $T$ and $\rho_N$ are given in units  MeV and fm$^{-3}$ respectively. }
\label{table_mass-shift}
\end{table}

\section{Summary}
\label{sec:5}

In this article, we have studied the effects of strong magnetic field on the in-medium mass of lowest charmonia, namely $J/\psi$ and $\eta_c$, in hot and dense isospin asymmetric nuclear matter using chiral $SU(3)$ model along with QCD sum rules. In effective chiral model, nucleons properties are modified through $\sigma$, $\zeta$, $\delta$, and $\chi$ fields resulting in the in-medium modifications of gluon condensates and masses of  $J/\psi$ and $\eta_c$. This interaction also accounts for scalar and number density of nucleons under the effect of magnetic field. For proton, since it has non-zero charge, we have contributions from the Landau levels hence it has strong influence, whereas for neutrons (zero charge) there are no Landau level contributions. Mainly, the in-medium modifications of $J/\psi$ and $\eta_c$ owes to the change in scalar gluon condensate and twist-2 gluon operator, which is obtained from the medium modification of scalar fields  $\sigma$, $\zeta$, $\delta$, and $\chi$. These fields  in strongly magnetized hot and dense isospin asymmetric (introduced by the $\delta$ and $\rho$ fields) nuclear matter is obtained by solving the coupled equations of motions (Eqs.(\ref{sigma}) to (\ref{chi})). We have observed that the effective mass of   $J/\psi$ and $\eta_c$ decrease with the increase in magnetic field except in the case of low-temperature regime of high isospin asymmetry, where increase in effective mass is observed due to the crossover behaviour of scalar fields and scalar densities of protons and neutrons. The mass modifications of these mesons are observed to be appreciable in high density and symmetric nuclear matter. It is also observed that the masses of  vector meson, $J/\psi$ and pseudoscalar meson, $\eta_c$ increase as we move towards high temperature. The observed mass-shift may be used to study  the decay of charmonia \cite{Friman2002} in the presence of magnetic field. The present work will shed light on non-perturbative effects of magnetic field on the properties of mesons and hadrons, which may help to understand the experimental observables such as decay width \cite{Friman2002,Golubeva2003} and  cross-section \cite{Khachatryan2017}, in asymmetric non-central heavy ion collision experiments. 

\begin{center}
 \section*{Acknowledgment}
\end{center}

One of the author, Rajesh Kumar sincerely acknowledge the support towards this work from Ministry of Science and Human Resources Development (MHRD), Government of India via Institute fellowship under National Institute of Technology Jalandhar.

\end{document}